\documentclass[preprint,review,12pt]{elsarticle}

\usepackage{float}
\usepackage{graphicx}
\usepackage{multirow}
\usepackage{pdflscape}
\usepackage{booktabs}
\usepackage{color,amsmath,amssymb}
\usepackage{color}
\graphicspath{{Figures/}}

\newcommand{\2}{$_{2}$}
\newcommand{\cm}{cm$^{-1}$}
\newcommand{\etal}{\textit{et al.}}
\newcommand{\abinitio}{\textit{ab initio}}

\begin{document}

\title{Room temperature line lists for CO\2 symmetric isotopologues with \textit{ab initio} computed intensities}

\author{Emil J. Zak$^a$, Jonathan Tennyson$^a$\footnote{To whom correspondence should be addressed; email: j.tennyson@ucl.ac.uk}, 
Oleg L. Polyansky$^a$, Lorenzo Lodi$^a$}

\author{Nikolay F. Zobov$^b$}

\author{Sergei A. Tashkun$^c$, Valery I. Perevalov$^c$}

\address{$^a$Department of Physics and Astronomy, University College London,\\
London, WC1E 6BT, UK}
\address{$^b$Institute of Applied Physics, Russian Academy of Sciences, Ulyanov Street 46, Nizhny Novgorod 603950, Russia } 
\date{\today}
\address{$^c$V.E. Zuev Institute of Atmospheric Optics, SB RAS, 1, Academician Zuev Square, Tomsk 634021, Russia}

\begin{abstract}
Remote sensing experiments require high-accuracy, preferably sub-percent, line intensities 
and in response to this need we present computed room temperature line lists for 
six symmetric isotopologues of carbon dioxide: $^{13}$C$^{16}$O$_2$, $^{14}$C$^{16}$O$_2$, 
$^{12}$C$^{17}$O$_2$, $^{12}$C$^{18}$O$_2$, $^{13}$C$^{17}$O$_2$ and 
$^{13}$C$^{18}$O$_2$, covering the range 0-8000 \cm.
Our calculation scheme is based on variational nuclear motion calculations and on a reliability analysis of the generated line intensities. Rotation-vibration wavefunctions and energy levels are computed using the DVR3D software suite and a high quality semi-empirical potential energy surface (PES), followed by computation of intensities using an  \abinitio\ dipole moment surface (DMS). 
Four line lists are computed for each isotopologue to quantify sensitivity to minor distortions of the PES/DMS. Reliable lines are benchmarked against recent state-of-the-art measurements and against the HITRAN2012 database, supporting the claim that the majority of line intensities for strong bands are predicted with sub-percent accuracy. Accurate line positions are generated using an effective Hamiltonian. We recommend the use of these line lists for future remote sensing studies and their inclusion in databases.
\end{abstract}

\maketitle
\newpage

\section{Introduction}

Remote sensing of carbon dioxide for atmospheric applications requires a 
detailed knowledge of the rotational-vibrational spectra of all its major 
isotopologues. It is commonly believed that such measurements should be 
supported by reference line intensities of 1$\%$ accuracy or better \cite{XOCO}. 
This is the main requirement for successful interpretation of data from the NASA 
Orbiting Carbon Observatory 2 (OCO-2) space mission, which is designed to 
monitor the concentration of carbon dioxide in Earth's atmosphere.
As was shown in our previous papers considering the main $^{12}$C$^{16}$O$_2$ 
isotopologue (denoted here as 626), a fully \abinitio\ dipole moment surface is 
capable of providing such a level of accuracy \cite{jt613,15ZaTePo.CO2}. The next 
step is to address other isotopologues, which are ubiquitous in natural samples 
and may interfere with spectral lines of the main isotopologue as well as other 
species \cite{13GaBaCa.CO2, 14KaCaMo.CO2, 15GeVaPh.CO2, 11GaPaLo.CO2}. Concentration 
measurements of trace compounds also require very accurate line positions, 
intensities and line profiles of several isotopologues at the same time.

Experimental line-intensity measurements for isotopologues other than the main 626 one are much more challenging due 
to their low natural abundance, which ranges from 1\% for $^{13}$C$^{16}$O$_2$ (636)
down to 10$^{-10}$ \% for the unstable $^{14}$C$^{16}$O$_2$ (646). Enriched samples 
feature the need for precise (\textit{a priori}) knowledge of isotopologue 
concentration. Therefore experimental accuracies of line intensities for less 
abundant isotopologues of carbon dioxide are in general lower than for the main 
isotopologue. A comprehensive source of spectroscopic data for carbon dioxide is 
the HITRAN database \cite{jt557}. 
The 2012 release of this database contains line lists for the 
$^{13}$C$^{16}$O$_2$ (636), $^{12}$C$^{17}$O$_2$ (727), $^{12}$C$^{18}$O$_2$ (828) 
and $^{13}$C$^{18}$O$_2$ (838) isotopologues, all featuring spectral gaps 
because of lack of experimental data. Line intensities included in the database 
have uncertainties of 5-20\%, thus well above current  requirements. The 
database comprises both semi-empirical (taken from the pre-release of the CDSD 
database \cite{08PeTa.CO2,15TaPeGa.CO2}) and purely experimental entries. 
Multiple data sources are reflected by discontinuities in intensity patterns of 
some bands \cite{15ZaTePo.CO2}.
Several measurements have been made since the last edition of the database, 
to pursue the elusive naturally abundant molecules, 
although these studies also have uncertainties well above the desired 1\% 
threshold. The most accurate recent experiments by Devi \etal\ 
\cite{16DeBeSu.CO2}, Jacquemart \etal\
\cite{12JaGuLy.CO2,14BoJaLy.CO2,15BoJaLy.CO2,15JaBoLy.CO2}, Karlovets \etal\ 
\cite{14KaCaMo.CO2} and Durry \etal\ \cite{10DuLiVi.CO2} address several bands 
of carbon dioxide isotopologues with unprecedented, though still unsatisfactory, 
intensity accuracies between 1\% and 20\%.

Among the few theoretical attempts to model high-resolution infrared spectra of 
carbon dioxide, the study by Huang \etal\ from the NASA Ames Research Center 
proved to be very accurate \cite{12HuScTa.CO2,14HuGaFr.CO2}. Line lists from 
Huang \etal\ provide both line positions and intensities covering the infrared and visible
spectral region ($J = 0 -150$) for room temperature (296~K) and 1000~K. 
Line positions, derived from a variational approach and based on semi-empirical 
potential energy surface (PES), are accurate to 0.03 -- 0.2 \cm. Although the 
Ames-1 line lists \cite{12HuScTa.CO2,14HuGaFr.CO2} show the best agreement with experiment among all variational 
calculations, semi-empirical approaches based on effective Hamiltonians can 
provide line positions with an accuracy at least one order of magnitude better 
\cite{08PeTa.CO2}. On the other hand, effective Hamiltonian models strongly 
depend on the quality of the input data, thus the accuracy and completeness of 
this technique are limited by experiment. 

A major advantage of the use of variational nuclear motion programs is that, 
within the limits of the Born-Oppenheimer approximation,
line intensities can be computed with the same accuracy for all
isotopolgues.
In the present study the computation 
of line intensities was based on \abinitio\ dipole moment surfaces (DMSs), and 
these were preliminarily tested by several authors 
\cite{15ZaTePo.CO2,15TaPeGa.CO2,16DeBeSu.CO2}. Some minor inaccuracies and 
discontinuities were discovered; however, comparisons showed overall good agreement 
with experiment.
A comprehensive literature review on both experimental and theoretical 
approaches to line positions and intensities is given in the recent work by 
Tashkun \etal\ \cite{15TaPeGa.CO2}. In this study the authors indicate and 
support the need for a unified theoretical treatment of line positions and 
intensities; whereas the former are largely covered by experiments facilitated 
by effective Hamiltonian models \cite{08PeTa.CO2}, the latter still await  high 
levels of accuracy, which we target in the present study. Due to lack of 
experimental data, a line list for the radioactive isotopologue $^{14}$C$^{16}$O$_2$ is not included in 
the HITRAN2012 and CDSD-296 databases. The Ames-1 line list is presently a 
unique source of high accuracy spectra for this species.

The unstable $^{14}$C$^{16}$O$_2$ isotopologue is of particular importance 
because of its usage in dating of biosamples and, more recently, in monitoring 
emissions, migrations and sinks of fossil fuel combustion products  
\cite{10ReRaPo.CO2, 10LeNaKr.CO2, 10RoSiGu.CO2} as well as for the assessment of 
contamination from nuclear power plants \cite{08ToWaTa.CO2}. Until recently, 
monitoring fossil fuel emission relied mostly on $\beta$-decay count 
measurements or mass spectrometry, both of which are high cost, invasive 
methods.

Despite its low natural atmospheric abundance, radiocarbon dioxide has been 
probed via optical spectroscopy methods \cite{15GeVaPh.CO2, 11GaPaLo.CO2, 
15GiGaMa.CO2, 11GaBaBo.CO2}. Recent advances in absorption laser spectroscopy 
provided an unprecedented tool for detection of species containing radiocarbon 
of ratios $^{14}$C/$^{12}$C down to parts per quadrillion. These measurements 
exploit saturated-absorption cavity ring down (SCAR) spectroscopy technique 
\cite{10GiBaBo.CO2} for the strongest fundamentals of $^{14}$CO$_2$ 
\cite{11GaPaLo.CO2,11GaBaBo.CO2}. The knowledge of accurate line 
intensities for several isotopologues at the same time is therefore a necessity 
for eliminating the unwanted noise sourced in traces of different isotopic 
carbon dioxide representatives. For instance the $P(20)$ line of the 00011 -- 
00001 band in 646, which is dedicated for radiocarbon measurements, above 
certain temperatures, heavily  interferes with the Lorentzian tail of the $P(19)$ 
line in the 05511 -- 05501 band of the 636 isotopologue \cite{13GaBaCa.CO2}. This 
raises difficulties in retrieving unbiased concentrations of the radioactive 
isotopologue. Similar problems occurred in measurements based on the 
$P(40)$ line of the $\nu_3$ band of $^{14}$CO$_2$ \cite{15McOgBe.CO2}.
In both cases accurate values of line intensities 
are required. Otherwise, as shown in \cite{15McOgBe.CO2}, calculation of the 
fraction of $^{14}$C in measured samples that employed a line strength taken 
from a theoretical approach, led to over 35 \% error in retrieved concentrations 
(as later confirmed by alternative experiments (AMS)). These observations were 
explained in terms of both inaccuracies of the line intensity and drawbacks of 
the spectroscopic fit model used, which fuels the need for reliable line 
intensity sources. Another successful technique further supporting this need 
was recently introduced by Genoud \etal\ \cite{15GeVaPh.CO2}, cavity ring-down spectroscopy with quantum cascade laser for monitoring of emissions from nuclear power plants. High quality line intensities for $^{12}$C$^{16}$O$_2$, $^{13}$C$^{16}$O$_2$ and $^{16}$O$^{12}$C$^{18}$O are also required for real-time detection methods based on quantum cascade lasers to monitor $^{13}$C/$^{12}$C isotope ratios in identification of bio-geo-chemical origins of carbon dioxide emissions from the soil-air interface \cite{16GuNuCh.CO2}. 
Spectra of isotopologues can be used for a variety of different tasks such as the recent suggestion that observations of absorptions by $^{13}$C$^{16}$O$_2$ in breath analysis provides a non-invasive means of diagnosing gastrointestinal cancers \cite{16KiShKo.CO2}.

In this work we aim to provide highly accurate line intensities in the 0 -- 8000 
\cm\ spectral region together with reliable semi-empirical line positions for 
six symmetric isotopologues of carbon dioxide; five naturally abundant: 636, 
727, 737, 828, 838 and one radioactive (646). An important advantage of a first 
principles theoretical approach is wide spectral coverage, in contrast to 
limited laser tuning capability of some measurements. For this reason databases like 
HITRAN, containing ro-vibrational spectra of small molecules, often utilize 
various experimental data sources, thereby giving up consistency of entries. 
This is not the case for the \abinitio\ approaches, which are believed to 
provide consistent accuracy of line intensities within a vibrational band 
\cite{jt522}.

Another issue related to post-processing of experimental data is the
functional form of the Herman-Wallis factors. These, as arbitrarily
chosen and empirically fitted, can become a source of unnatural biases
for high $J$ transitions \cite{15ZaTePo.CO2}. The theoretical model
implemented here should be inherently free from this problem, as the
rotational contribution to line intensities is computed directly. The
only underlying source of errors within the Born-Oppenheimer
approximation is potential energy surface (PES) and dipole moment
surface (DMS). The quality of both surfaces is assessed here by
employing the line sensitivity analysis procedure introduced by Lodi
and Tennyson \cite{jt522}, which requires computation of at least four
line lists for each isotopologue. This allows us to detect resonances
\cite{15ZaTePo.CO2}, that affect line intensities and significantly
diminish the reliability of data provided.

The final results are given in the form of line lists with associated
uncertainties, which are available in the supplementary materials.
Uniformity of errors (except for the regions affected by resonance
interactions) accompanied by the precise reproduction of observed line
intensities at the sub-percent level for the majority of strong bands
\cite{15ZaTePo.CO2} suggests our results provide a viable update to
the current, 2012 version of the HITRAN database.

\section{Methodology}

The Lodi-Tennyson methodology presented in detail elsewhere \cite{15ZaTePo.CO2,jt522} 
is used to validate line lists on a purely theoretical basis. For 
each isotopologue four line lists were computed, based on set of two PESs and 
two DMSs. These surfaces come from two sources: the semi-empirical Ames-1 PES 
and DMS from Huang \etal\ \cite{12HuScTa.CO2}, an \abinitio\ PES and DMS (UCL 
DMS) computed by us, where the former one was subsequently fitted to observed $J = 0-2$ levels 
 (called below: Fitted PES). Details of all surfaces 
were presented before \cite{15ZaTePo.CO2}. These four high-quality surfaces are 
used in nuclear motion calculations to obtain rotational-vibrational energy 
levels, wavefunctions and line intensities.

\subsection{Nuclear motion calculations}

The strategy for solving the nuclear-motion problem employed here is analogous 
to the one presented in \cite{15ZaTePo.CO2}. In the Born-Oppenheimer 
approximation the PES and DMS are 
isotopologue independent. Therefore the only parameters that distinguish between 
different isotopologues are the nuclear masses entering the expression for the 
kinetic energy operator (KEO) of the nuclei. Our approach is based on an exact,
within the Born-Oppenheimer approximation, 
KEO which is used in a two step procedure \cite{jt48,jt66,jt96} of solving the 
nuclear Schr\"odinger equation. The first step involves solving the Coriolis-decoupled ro-vibrational 
motion problem for every $(J,|k|)$ quantum number combination separately to supply a basis for the 
second step, where the full ro-vibrational Hamiltonian is considered. 
The DVR technique implemented in the DVR3D suite \cite{jt338} is used to 
represent the Hamiltonian matrix. Diagonalisation of this matrix  provides 
rotational-vibrational energy levels and wavefunctions. In the present study we 
used symmetrized Radau coordinates and a bisector embedding to define nuclear 
degrees of freedom in the body-fixed coordinate system.

Nuclear masses for isotopes of carbon and oxygen were used: 11.996709 Da ($^{12}$C), 
13.000439 Da ($^{13}$C), 14.000327 Da ($^{14}$C), 15.990525 Da ($^{16}$O), 
16.995245 Da ($^{17}$O) and 17.995275 Da ($^{18}$O) \cite{95AuWa.CO2}.  In the first step of the 
computation (program DVR3DJZ) the  Born-Oppenheimer ro-vibrational wavefunctions 
were expanded in Morse-like oscillator basis functions for stretching 
coordinates and Legendre polynomials for the bending one; symmetrised Raudau 
internal coordinates were used. The converged DVR basis set associated with 
Gauss-Legendre quadrature points contained 30 radial and 120 angular functions, 
respectively. In the second step (program ROTLEV3b) we employed a $J-$dependent 
basis set of symmetry-adapted symmetric-top rotational functions. The $J$ ranges considered
were chosen to match those present in the HITRAN2012 database (see Table 
\ref{table:info}). The same set of parameters was used to evaluate 
ro-vibrational energies computed from Ames-1 and fitted PESs. Calculation of 
line strengths used the DIPOLE program and require both DMS and ro-vibrational 
wavefunctions as input functions. Transformation from line strengths to 
transition intensities included scaling by natural isotopic abundance and 
multiplication by the appropriate spin statistical weights. Partition functions 
at 296~K were taken from Huang \etal\ \cite{12HuScTa.CO2} and they agree to 
better than 0.1\% with partition functions from the present computation based on 
Ames-1 PES (see Table \ref{table:info}). A natural-abundance weighted intensity cut-off was then set to 
$10^{-30} $ cm/molecule. In the case of the radioactive isotopologue (646) we 
assumed unit abundance and increased the cut-off value to $10^{-27}$ 
cm/molecule.

\subsection{Estimation of the intensity uncertainties}

The {\it ab initio} DMS is a primary source of inaccuracies in line
intensities.  The accuracy of the UCL DMS has been proven to be at
sub-percent level for several strong bands (stronger than $10^{-23}$
cm/molecule) below 8000 \cm \cite{jt613,15ZaTePo.CO2}. A key feature
of using an {\it ab initio} DMS with a variational nuclear motion
calculation is that entire vibrational bands are reproduced with
similar accuracy.  The reliability of line intensities obtained
theoretically is correlated with the quality of $J$-dependent
ro-vibrational wavefunctions, hence with an underlying PES.
Wavefunctions play an important role in capturing the interaction
between different vibrational states. Such resonance interactions can
lead to intensity stealing and, particularly for so-called dark
states, huge changes in transition intensities.

Indeed, the sensitivity analysis performed for 626 \cite{15ZaTePo.CO2}
indicated that certain ro-vibrational energy levels of CO$_2$ are
vulnerable to interactions induced by Coriolis-type terms in the
Hamiltonian. These resonances may change intensities of lines,
particularly for weak transitions, by several orders of magnitude.
This effect manifests itself especially when the perturbing band is
strong. For this reason even minor inaccuracies in the PES in regions
where such interactions occur can significantly change the intensity
of lines, hence not reflecting the accuracy of the DMS. Thus, it is
important to filter out bands or specific transitions with perturbed
intensities which should not be trusted. A few such bands have
been reported for the main isotopologue of CO$_2$ (626)
\cite{15ZaTePo.CO2,13HuFrTa.CO2} and other isotopologues
\cite{14KaCaMo.CO2,15BoJaLy.CO2,15JaBoLy.CO2}.

Here we adopt a scheme from our previous works
\citep{15ZaTePo.CO2,jt522}, where a method of locating resonances was
developed and tested on H$_2$O and $^{12}$C$^{16}$O$_2$.
Cross-comparison of four line lists yielded an intensity scatter
factor $\rho$ assigned to every line, which is defined as the ratio of
the strongest to the weakest transition in a set of four intensities
matching the same transition line. The four line lists read: Ames-1
PES and Ames DMS (named AA), Ames-1 PES and UCL DMS (named AU), Fitted
PES and Ames DMS (named FA), Fitted PES and UCL DMS (named FU). Our
base model, which should on theoretical grounds be the most accurate,
is AU; it is from this model that our underlying uncertainties are drawn. For transitions involved in resonance interactions,
calculations with different procedures should give markedly different
results. The trustworthy line intensities should be stable under minor
PES/DMS modifications, hence should feature a small scatter factor.
Based on the statistics of the scatter factor we established arbitrary
limits on $\rho$ for a line to be considered stable ($1.0 \leq
\rho<2.5$), intermediate ($2.5 \leq \rho<4.0$) and unstable ($\rho
\geq 4.0$). These values are the same for all isotopologues.  We
believe that this descriptor gives a robust measure of sensitivity of
line intensities to small PES changes, and hence reflects the
reliability of a theoretically evaluated line intensity.

\subsection{Line positions}\label{line_positions}

As discussed above line positions for the “recommended UCL-IAO 
line list” are taken from the CDSD-296 database \cite{15TaPeGa.CO2}. These line 
positions were calculated within the framework of the effective Hamiltonian 
approach for which the partly-reduced polyad model was used 
\cite{92TeSuPe.CO2,98TaPeTe.CO2,00TaPeTe.CO2}. This model takes into account the 
resonance anharmonic, anharmonic+\textit{l}-type and Coriolis interactions. The 
global weighted fits of the effective Hamiltonian parameters to the observed 
line positions collected from the literature were performed. The observations 
cover the 0.7 - 13~633 \cm\ range, but for the rare isotopologues the ranges of 
the observations are considerably smaller. A review of the observed line 
positions used is given elsewhere \cite{15TaPeGa.CO2}. Each isotopologue was considered 
separately. In the case of isotopologues for which the observed line positions 
are scarce the effective Hamiltonian parameters obtained from the 
multi-isotopologue fitting \cite{12Ta.CO2} were used as an initial guess for further 
refinement of some of the parameters by separate fitting. The fitted sets of the 
effective Hamiltonian parameters reproduce the observed line positions 
practically within their experimental uncertainties, from 10$^{-9}$ \cm\ to 0.01 
\cm. The good predictive ability of the resulting effective Hamiltonian 
parameters has been demonstrated many times \cite{14KaCaMo.CO2,12JaGuLy.CO2,14BoJaLy.CO2,15BoJaLy.CO2,13KaCaMo.CO2,14KaCaMoII.CO2}. 
The exceptions are several bands of the asymmetric 
isotopologues which are perturbed by the interpolyad resonance anharmonic 
interactions. Some of the examples are given in 
Refs.\cite{14KaCaMo.CO2,13KaCaMo.CO2,14KaCaMoII.CO2}. Usually these  
perturbed bands are very weak.

\subsection{Pseudo-experimental refinements}

Energy levels taken from the effective Hamiltonian (EH) are generally considered to be accurate to 0.002 
\cm\ or better, which is usually more 
accurate than those given by PES based studies \cite{15TaPeGa.CO2}. Multi-isotopologue fits 
allowed for completeness of EH line lists even for less abundant isotopologues 
\cite{15KaPe.CO2}. From this reason our recommended line lists include line 
positions from the Effective Hamiltonian calculations provided by the CDSD-296 
database \cite{15TaPeGa.CO2}. However, the very limited experimental data for 
$^{14}$C$^{16}$O$_2$ did not allow us to extract EH parameters for this 
isotopologue. 
Therefore an alternative source of line positions is needed here.

Comparison of DVR3D (AMES-1 PES) line positions with EH values for the 
$^{13}$C$^{16}$O$_2$ isotopologue shows that on average agreement between the 
two is at the 0.02 \cm\ level.  Within one vibrational band the residuals 
between EH and DVR entries remain almost constant. This observation suggests the 
viability of correcting the DVR3D line positions with respective differences for 
the main isotopologue. Such an approach has recently been shown to give
excellent results for H$_2$$^{18}$O and H$_2$$^{17}$O \cite{jt522}.

Thus, an attempt to correct DVR3D line positions with the 
effective Hamiltonian values was made. First, energy differences between 
corresponding EH and DVR energy levels for the main isotopologue were taken: 
$\Delta E(626)=E^{626}_{EH}-E^{626}_{DVR}$. Next, each energy level of a given 
less abundant isotopologue was refined by adding respective difference to the 
DVR-computed value: $E^{ISO}=E^{ISO}_{DVR}+\Delta E(626)$. These refined energy 
levels were then compared to EH values.  
Application of the above procedure to symmetric isotopologues of CO$_2$ resulted 
however in increased deviations between DVR and EH energy levels, hence should 
be considered here as unsuccessful. The reason for this is not entirely
clear but the cited study for water \cite{jt522} took considerable
care over the corrections to the Born-Oppenheimer approximation while
the results here were based on a PES simultaneously fitted to data
from several isotopologues with no allowance for the beyond 
Born-Oppenheimer effects.
For this reason pure DVR3D line positions 
were used in the recommended line list for $^{14}$C$^{16}$O$_2$.

\section{Results and Discussion}

Below we present an overview of our line lists. The next subsection contains 
statistical analysis of the scatter factor for each isotopologue, which will be 
further used in detecting resonances in the following subsection. After that, we 
compare our present results to recent highly accurate measurements, which were not 
included in the latest release of the HITRAN database. Next, a detailed 
comparison with the HITRAN2012 and CDSD-296 databases is performed. Finally we 
discuss the radioactive 646 isotopologue in the environmental context, as one of 
possible applications of present results. 

\subsection{Overview}

Table \ref{table:info} contains general information about our line lists. 

\begin{table}[H]
\setlength{\tabcolsep}{2pt}
\caption{General information on computed room temperature ($T$=296~K) line lists.}
\vspace{0.2cm}
\footnotesize
\begin{tabular}{l l l l l l l}
\hline\hline
Isotopologue			 & 	636 	& 646		 & 727 		& 737 		& 828 		& 838 \\ [0.3ex] 
\hline 
ZPE$^a$ [\cm] 			& 2483.08	& 2436.75	& 2500.75  	 & 2447.50 	& 2469.05 	& 2415.39  	\\
J$_{MAX}$				& 119 		& 130 		& 99   		 & 50 		& 101 		& 50 		\\
Spin factors (ortho:para)	& 2:0 	 	& 7:0 		&  15:21 	 & 30:42  	&  1:0 		&  2:0		\\
Q$_{296}$(This work)	&  576.652 	& 2033.395   &  10~902.24   & 21~758.08 	& 323.438 	& 644.754  	\\
Q$_{296}$(CDSD-296)$^b$	&  576.652	& N/A   	&  10~971.90 & 22~129.96  & 323.418 	& 652.234  	\\
Q$_{296}$(Ames-296)$^c$	&  576.644 	& 2033.353   &  10~971.91 & 22~129.96  & 323.424 	& 652.242	\\
Q$_{296}$(HITRAN)$^d$ 	&  578.408  & N/A 		&  11~001.67 & N/A  	&  324.211 	& 653.756 	\\

Abundance $^e$ & \scriptsize $0.011057$ & \scriptsize $1.0$& \scriptsize $0.13685\times10^{-6}$ & \scriptsize $0.15375\times10^{-8}$ & \scriptsize  $0.39556\times10^{-5}$ &  \scriptsize $0.44440\times10^{-7}$\\
\footnotesize
No. lines (This work) $^f$ 	& 68~635 	&   41~610	&	6530	& 	1501	& 	10~441	& 2637		\\
No. lines (CDSD-296) $^f$ 	& 68~640 	&   N/A		&	6530	& 	1500	& 	10~444	& 2635		\\
No. lines (Ames-296) $^f$ 	& 68~739 	&   42~072	&	6545	& 	1634	& 	10~531	& 3050		\\
No. HITRAN lines 		& 68~856 	&	N/A		&  	5187	& 	N/A		& 	7070	& 	121		\\
No. HITRAN matches$^g$	& 68~856 	&	N/A		&   5187	& 	N/A		& 	7069	&	121		\\
\hline\hline
\end{tabular}
\label{table:info}
$^a$ Zero point energy based on the Ames-1 PES; 
$^b$ 2015 Edition of CDSD \cite{15TaPeGa.CO2}; 
$^c$ \cite{12HuScTa.CO2}; 
$^d$ TIPS-2011 \cite{11LaGaLa.CO2};  
$^e$  HITRAN2012 abundances were taken from Ref. \cite{12HuScTa.CO2}; 
$^f$ For $10^{-27}$ cm/molecule intensity cut-off for the 646 isotopologues and $10^{-30}$ cm/molecule
after scaling for natural abundance for the other isotopologues;   
$^g$ UCL-IAO line list with $10^{-33}$ cm/molecule intensity cut-off was used in the comparison.
\end{table}

Partition functions  in this work  were calculated using Ames-1 PES, and these 
are compared to HITRAN2012, Ames and CDSD-296 line lists. We can see from the 
Table \ref{table:info} that for isotopologues: 636, 646 and 828 the agreement 
between Ames work and our values is at $0.002 \%$ level, as expected from runs 
based on the same PES. Partition functions calculated by us should however be 
treated as provisional for the 737 and the 838 isotopologue, as $J$-range 
employed here did not allow for full convergence of the partition function. On 
average, the computed partition functions agree excellent with CDSD-296, and are 
systematically shifted by -$0.3 \%$ with respect to HITRAN2012. This fact must 
be taken into account in comparisons aimed at sub-percent accuracy of line 
intensities. For line intensity calculations we used Ames partition functions 
from Huang \etal\ \cite{12HuScTa.CO2}. The HITRAN2012 database has significant spectral gaps for 
less abundant isotopologues, which are all covered by our line lists. Lines 
present in HITRAN2012 were completely matched to our lines using the  energy 
level comparison technique.

\subsection{Scatter factor statistics}

\begin{figure}[H]
  \includegraphics[width=14cm]{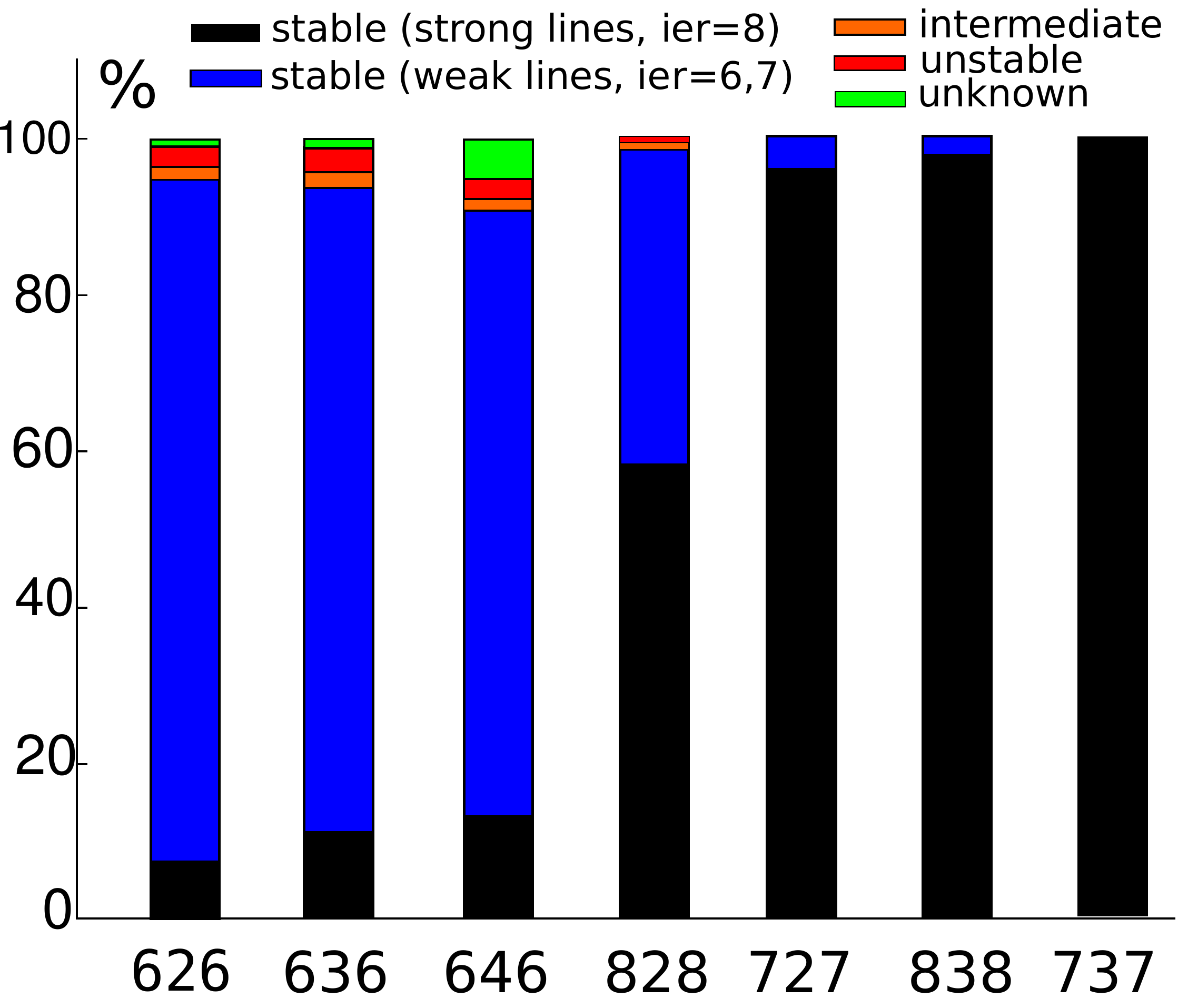}
\caption{Scatter factor statistics for six isotopologues of carbon dioxide. 
Respective colours denote percentages of lines classified to particular stability domain. The $y$ axis corresponds to percentage of total lines present in our line list. Black regions give percentage of stable and strong lines ($>10^{-23}$ cm/molecule), for which the highest HITRAN intensity accuracy code was assigned ($ier=8$).}
\centering
\label{fig:rho_stat}
\end{figure}

Figure \ref{fig:rho_stat} gives an overview of the scatter factor statistics for 
all isotopologues considered. It is readily seen that the majority of lines are 
classified as stable, that is have scatter factor less than 2.5 (marked in blue 
in Figure \ref{fig:rho_stat}). For $^{13}$C$^{17}$O$_2$ and $^{13}$C$^{18}$O$_2$ 
all lines are stable. Line lists for $^{13}$C$^{16}$O$_2$ and 
$^{14}$C$^{16}$O$_2$ have 3.5\% and 3.3\% of unstable lines respectively, 
whereas the $^{12}$C$^{18}$O$_2$ line list features only 0.5\% of unstable 
lines. Whenever a line-to-line match between a Fitted PES derived line and an 
Ames-1 based line was 
impossible by our automatic procedure due to ambiguity in matching of energy 
levels, manual attempts were made. Nevertheless a small fraction of lines 
remained unmatched, resulting in 'unknown' scatter factors; the numbers of such 
unmatched lines is small for all isotopologues. The blue dashed area denotes 
stable and strong lines ($>10^{-23}$ cm/molecule), for which the most accurate 
HITRAN accuracy index was assigned ($ier = 8$). The HITRAN uncertainty index 
value ($ier$) equal to 8 corresponds to accuracy of line intensity better than 
1\%. Consequently, $ier = 7$ stands for 1 -- 2\% accuracy, $ier = 6$ stands for 
2 -- 5\% accuracy, $ier = 5$ stands for 5 -- 10\% accuracy, $ier = 4$ stands for 
10 -- 20\% accuracy and $ier = 3$ stands for line intensities accurate to 20\% 
or worse.

\subsection{Resonances}

\begin{figure}[H]
  \includegraphics[width=14cm]{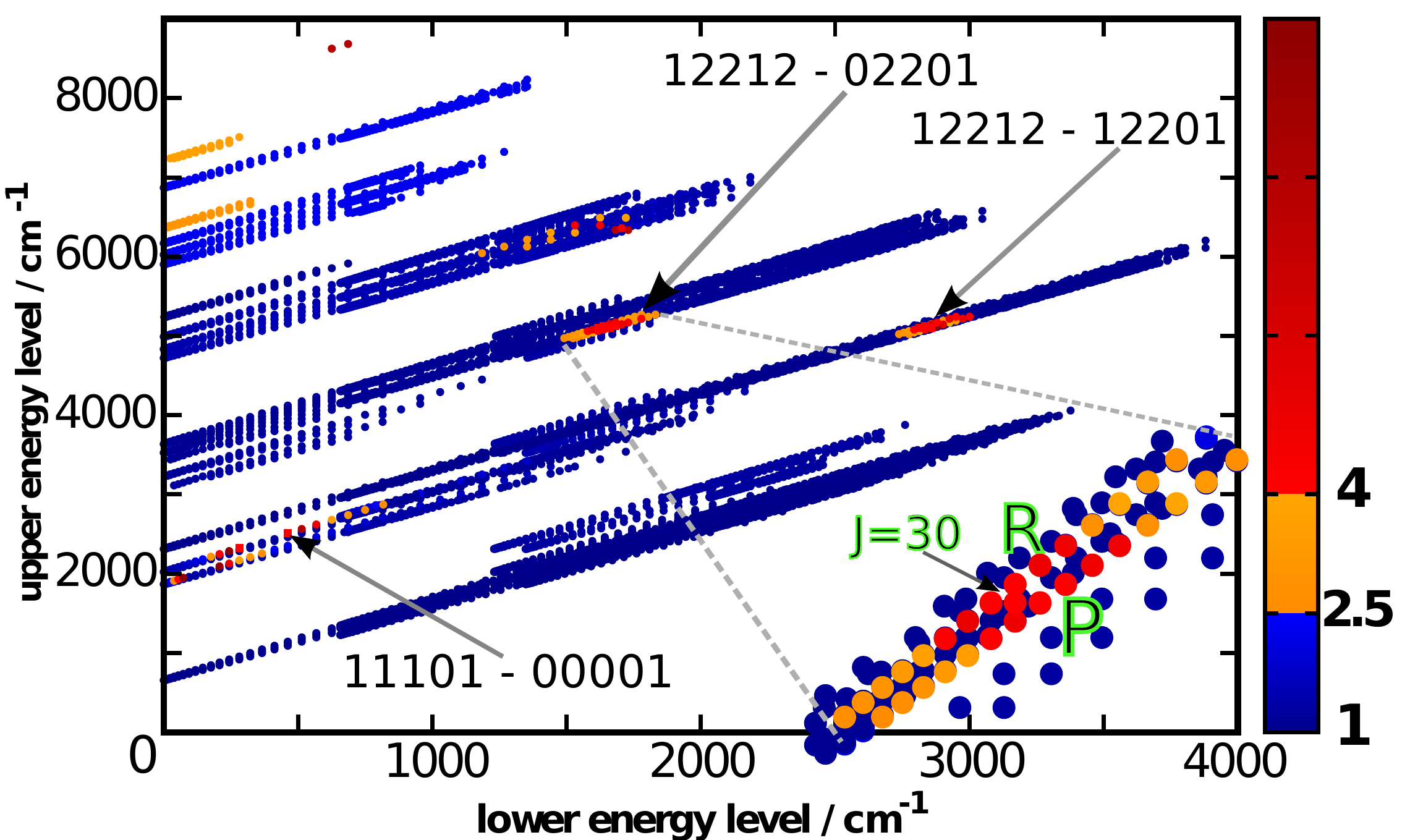}
\caption{Scatter factor map for the 828 isotopologue. Colour coding denotes respective classification of lines: blue  stands for stable lines, orange for intermediate lines and red for unstable lines. The arrows indicate selected bands for which a $J$-localized peak in the scatter factor is observed. The  zoomed inset in right bottom corner shows the peak region of the scatter factor for the 12212 -- 02201 band. Both P and R branches are affected by the interaction around $J=30$. }
\centering
\label{fig:map828}
\end{figure}

Detailed information on the energetic distribution of the scatter factor may be 
extracted from a 'transition stability map', which is a useful tool in searching for
resonances, as exemplified in Figure \ref{fig:map828}, which presents an example 
of such a map for the 828 isotopologue. The advantage of this particular 
representation is that one gains a full overview of all energetic regions, where 
transition intensities appear to be sensitive to minor inaccuracies of the PES. 
These lines are marked as red dots in Figure \ref{fig:map828}.

Figure \ref{fig:map828} also illustrates the general trend of
decreasing stability of lines with increasing energy of states
involved in a transition. This has been already observed for the main
626 isotopologue in \cite{15ZaTePo.CO2}. Sporadic red points localized
in small energetic areas are indicative of $J$-localized resonances,
while long chains of unstable points suggest instability of whole
bands. It is instructive to give a more detailed insight into
resonances by plotting scatter factor as a function of $m$ quantum
number for each band separately within a given polyad number change
($\Delta P$).  The $m$ quantum number is defined as equal to -J(lower
energy level) for the P branch, J(lower energy level) for the Q
branch, and J(lower energy level)+1 for the R branch.  The polyad
number for carbon dioxide is defined as $P=2\nu_1+\nu_2+3\nu_3$, where
$\nu_1, \nu_2, \nu_3$ are the vibrational quantum numbers of symmetric
stretching, bending and asymmetric stretching, respectively.

\begin{figure}[H]
  \includegraphics[width=14cm]{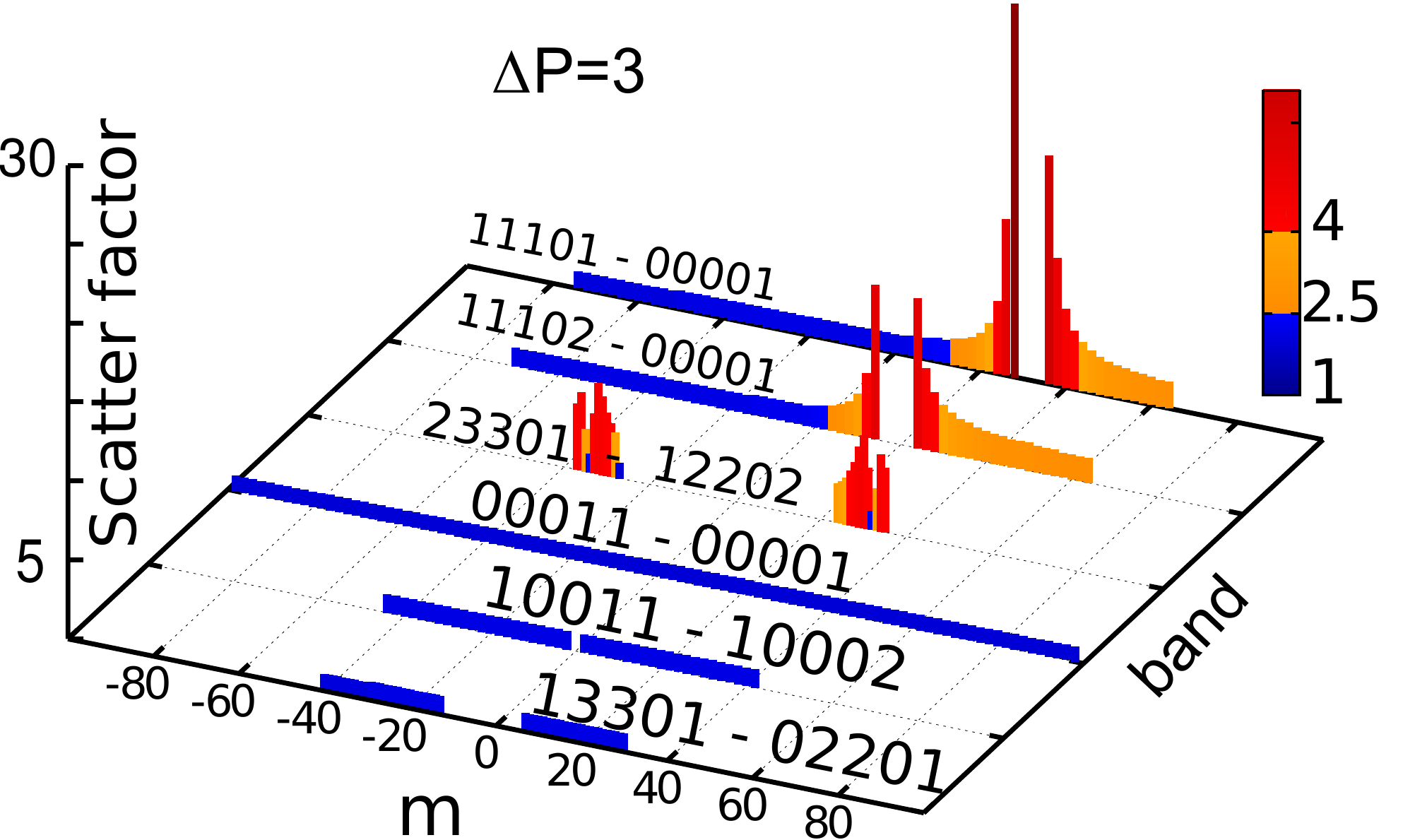}
\caption{Scatter factor distribution for selected bands of 828 with polyad change $\Delta P = 3$. Color code denotes classification of transition as stable (blue), orange (intermediate) or unstable (red), measured by the scatter factor.}
\centering
\label{fig:828_P3_4d}
\end{figure}

\begin{figure}[H]
  \includegraphics[width=14cm]{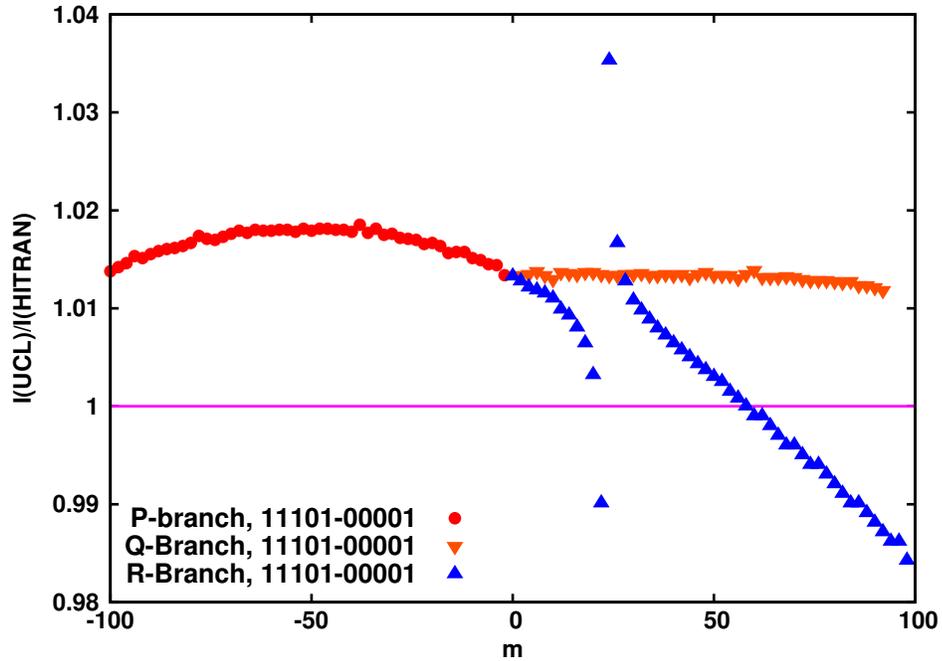}
\caption{Relative intensities  plotted against HITRAN2012 line 
intensities for 11101--00001 band for the 636 isotopologue. This is an example 
of a band involved in resonant Coriolis interaction.}
\centering
\label{fig:636_11101}
\end{figure}

\begin{figure}[H]
  \includegraphics[width=14cm]{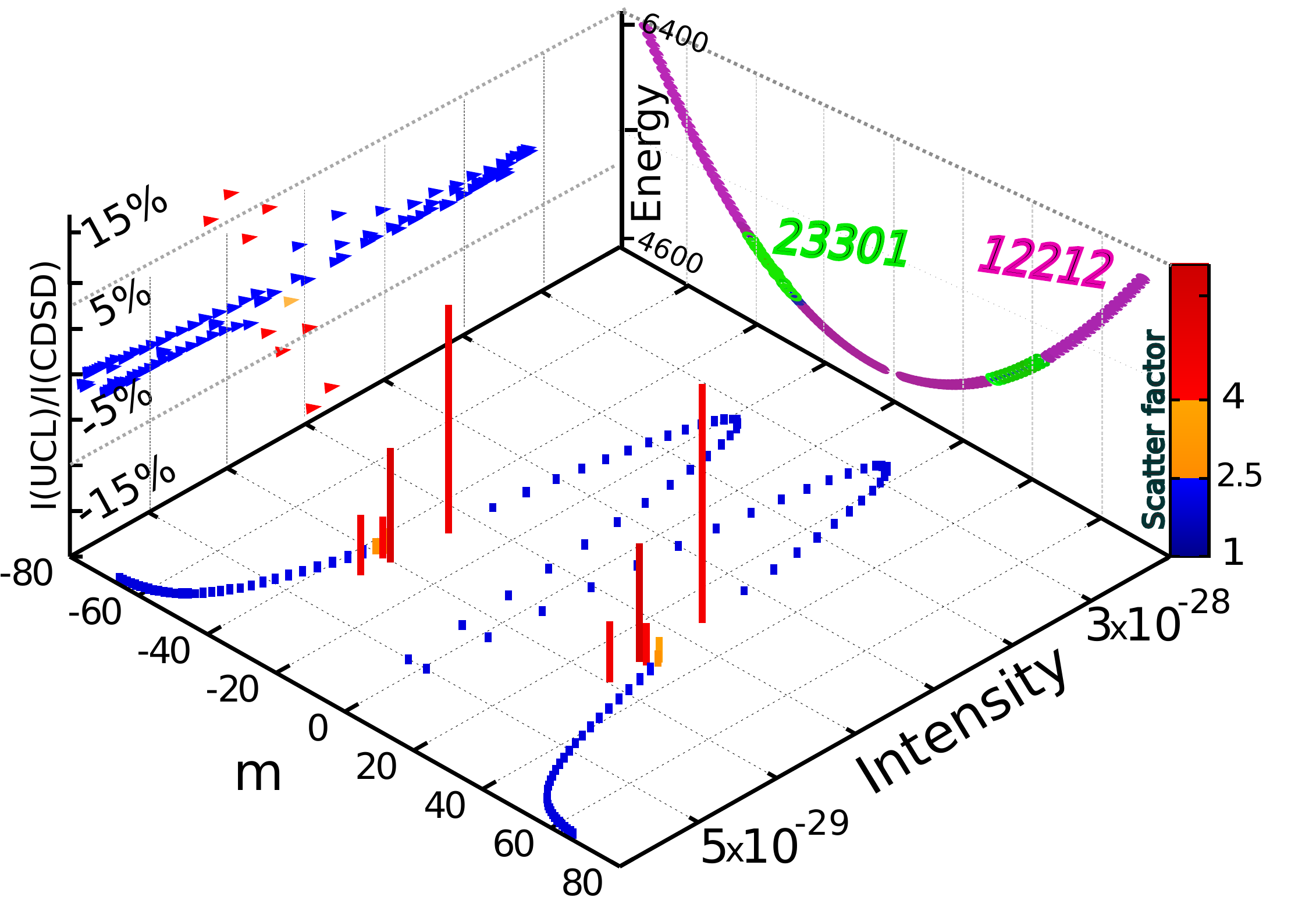}
\caption{Multidimensional graph characterising the 12212 -- 02201 band of 
$^{12}$C$^{18}$O$_2$. The base plane depicts $m$ dependence of line intensities 
with bar height and color code measuring the value of the scatter factor. The 
far right plane represents $m$ dependence of energy levels of the perturbed 
state (12212) and perturber (23301), which happen to nearly overlap around 
$m=\pm 36$. Left plane gives intensity ratios of lines taken from UCL-IAO line 
list and CDSD-296 database. }
\centering
\label{fig:12212-23301}
\end{figure}

In general bands with polyad number change equal to 1 ($\Delta P = 1$)
are stable. For $\Delta P = 3$ three unstable bands were found: 23301
-- 12202 , 11101 -- 00001 and 11102 -- 00001. The first of the three
bands contain transitions for which upper energy levels (localized
around a particular $J$ value) become energetically close to
rotational states of some other vibrational state; in this case to
levels from the 12212 state. This may lead to a strong resonance
interaction between states. In the case of the last two bands, an
intensity borrowing mechanism from the strong asymmetric stretching
fundamental is responsible for the instability of line intensities
around a particular $J$ (see Figure \ref{fig:828_P3_4d} and Figure
\ref{fig:636_11101}).  For $\Delta P = 5$, both 12212 -- 02201 and
23301 -- 02201 bands are subject to a $J$-localized resonance, as
depicted in Figure \ref{fig:12212-23301}. This is due to mutual
interaction of the upper levels of these bands, which are
energetically close. For $\Delta P = 7$, the 22213 -- 02201 band
exhibits a weak $J$-localised peak in the scatter factor around
$J=34$. The 31101 -- 00001 band is weakly perturbed by interaction
between the 31101 and 20012 states in the vicinity of $J=68$. Bands
with higher polyad change number ($\Delta P = 9,11$) are in general
less stable, following uniform distribution of the scatter factor.

\begin{figure}[H]
  \includegraphics[width=14cm]{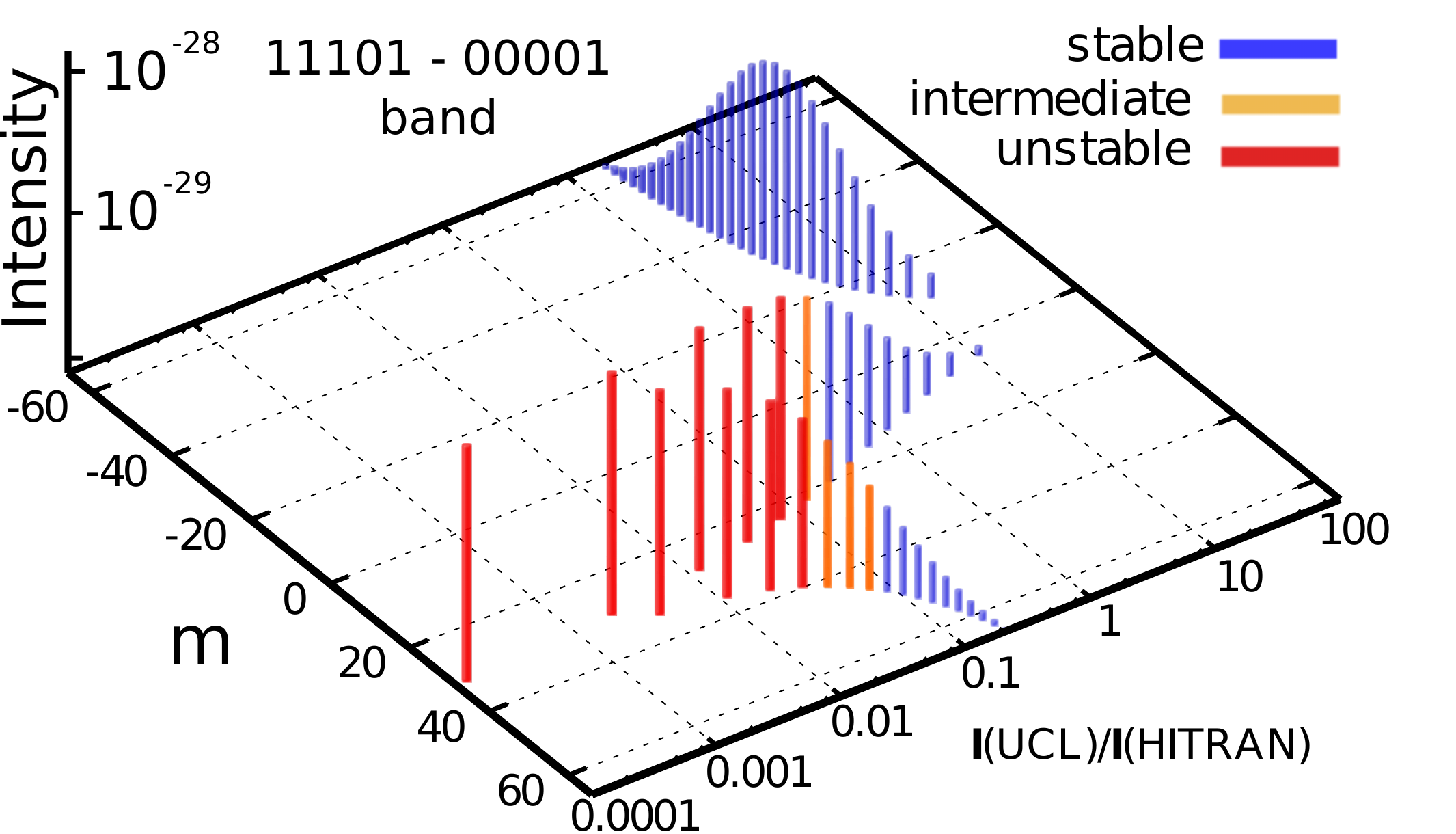}
\caption{Relative intensities plotted against HITRAN2012 line intensities for the 
11101--00001 band for the 828 isotopologue. This is an example of a band 
involved in resonant Coriolis interaction. Blue, orange and red points denote 
stable, intermediate and unstable lines, respectively.}
\centering
\label{fig:11101}
\end{figure}

Resonances occur when ro-vibrational energy levels of two or more states cross 
or nearly cross in the vicinity of a single $J$ value. A prominent example of near 
crossing situation is the 11101 -- 00001 band, which is perturbed by the 00011 
state (intrapolyad interaction). Because the 00011 -- 00001 fundamental is very 
strong and the perturbed band is relatively weak, significant intensity stealing 
is observed. This case is depicted in Figure \ref{fig:11101}. The relative intensity 
representation supplied by Figure \ref{fig:11101} leads to the immediate conclusion of 
unreliability on the theoretical line intensities of transitions affected by 
resonances. 

In Figure \ref{fig:11101} colour coding shows the stability of the transition 
intensity. A J-localized resonance is visible around $m=+36$, clearly 
correlating with both high instability of lines (marked by red points) and large 
deviations from HITRAN2012 line intensities. 

This quasi-singularity in line intensity occurs due to Coriolis interaction with 
the strong 00011--00001 band, which equally perturbs P and R branches of the 
11101--00001 band, and manifests itself by intensity borrowing, which in turn 
leads to the strengthening of the P-branch and to suppression of the R-branch. 
This observation confirms our previous predictions for existence of such 
perturbation in the main isotopologue \cite{15ZaTePo.CO2}. For this reason, in 
the final recommended line list we replace line intensities perturbed by these
Coriolis interactions with semi-empirical data from the CDSD-296 database. 

A view of the 636 isotopologue in Figure \ref{fig:636_11101} supports this 
thesis. Similar behaviour is observed for other isotopologues. The 636 case 
clearly shows that this type of interaction is branch-specific and $J$-specific 
as illustrated in Figure \ref{fig:636_11101}.

Another example of the intrapolyad interaction is the pair: 23301 (perturber) and 
12212 -- 02201 (perturbed band), for which we depict the intensities scheme in 
Figure \ref{fig:12212-23301}.

Figure \ref{fig:12212-23301} shows perfect correlation between line stability 
measured by the scatter factor and agreement with CDSD-296 line intensities, 
where large discrepancies surround the region of elevated scatter factor (marked 
with red filled triangles in Figure \ref{fig:12212-23301}). Very similar 
behaviour for line positions of the 12212 -- 02201 band  was noted by Borkov 
\etal\ for 727 \cite{15BoJaLy.CO2}, whose simple polynomial fit of the line 
positions resulted in a $J$-localised quasi-singularity in deviation of line 
positions. In such cases, effective Hamiltonian calculations have proved to be 
very successful and should be trusted regardless of resonances 
\cite{14KaCaMo.CO2, 15JaBoLy.CO2}.

One would expect that at least some of the large deviations in line
intensities (see Figures \ref{fig:636_11101} and
\ref{fig:12212-23301}) can be assigned to the influence of a
resonance. Indeed, the correlation between high deviations in
intensity and high scatter factor values is strikingly high, and
reliability of the data provided might be unresolvable in favor of
either approach.  Nevertheless, as many HITRAN2012 transitions comes
from CDSD-296 semi-empirical calculations, one should expect them to
be trustworthy.  This behavior is common for all bands in our line
lists. Therefore we may consider the scatter factors used as a
legitimate measure of reliability of a theoretical line list.

In the recommended line lists given in the supplementary materials, we
replace the intensities of lines identified as unstable with entries
from the effective Hamiltonian calculations, which for these cases are
much more reliable, being experimentally based. Line lists with
\abinitio\ intensities can be found in supplementary material in files
named as 'UCL-IAO-296-isotopologue\_ name.dat'.

\subsection{Comparison with recent measurements}

Numerous experimental line intensity data sources are available throughout 
the literature (see refs. \cite{15TaPeGa.CO2,11TaPe.CO2} for a comprehensive 
review), of which only few have high precision. Our attention is therefore 
focused on the most recent and the most accurate measurements. 
We assume that up to 2011 all highly accurate measurements and 
computations are captured by the 2012 edition of HITRAN, which will be analysed 
below. 

A notable advantage of our \abinitio\ approach is constant accuracy of intensity 
of entire bands, which is also believed to be transferable between different 
isotopologues. This is because all line intensities are computed using the same 
PES and DMS. Assuming that non-Born-Oppenheimer corrections are negligible we 
should expect an accuracy of 1\% or better for strong and stable bands of all 
symmetric isotopologues. These bands are defined as ones containing
transitions stronger than $10^{-23}$ cm/molecule at 100\%\ abundance.   For the weak stable and intermediate bands
this accuracy reduces to 1 -- 3\%. Such 
levels of accuracy are generally beyond reach of current experimental 
techniques, especially for the less abundant species.

\subsection{Isotopologue 636}

\begin{figure}[H]
  \includegraphics[width=14cm]{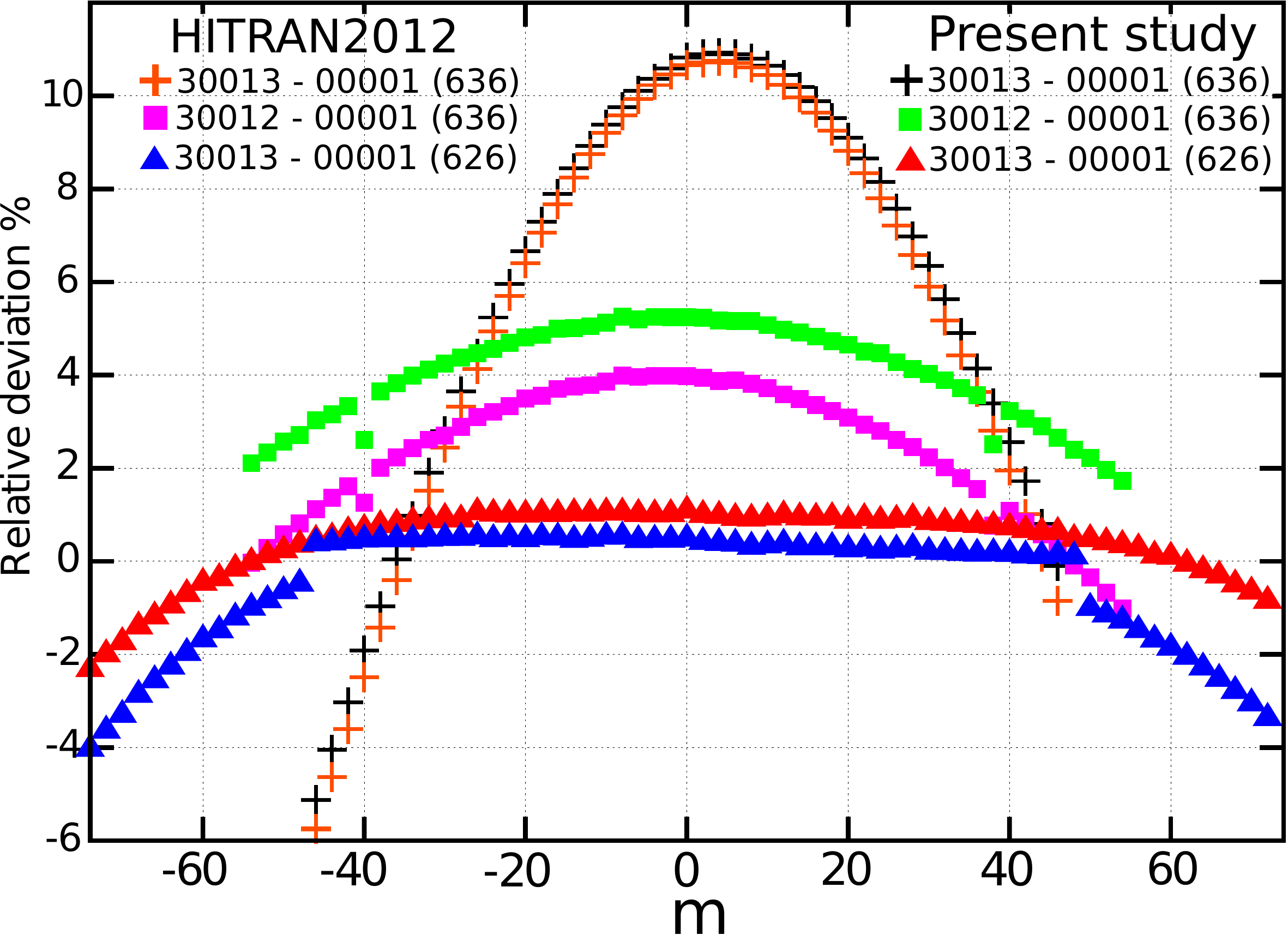}
\caption{Relative deviations (PS/Experiment) of line intensities from measurements by Devi \etal\ \cite{16DeBeSu.CO2} plotted against $m$ quantum number. Blue and red triangles denote the 300013 -- 00001 band of the 626 isotopologue taken from HITRAN2012 and present study, respectively. Purple and green squares stand for line intensities of the 30012 -- 00001 band of the 636 isotopologue taken from HITRAN2012 and present study, respectively. Orange and grey circles give the line intensities of the 30013 -- 00001 band of the 636 isotopologue taken from HITRAN2012 and present study, respectively. Zero relative deviation means 100\% agreement with Devi \etal\ }
\centering
\label{fig:Devi}
\end{figure}

Recently Devi \etal\ \cite{16DeBeSu.CO2} performed precise measurements of line 
intensities of the 626, 636 and 628 isotopologues of carbon dioxide in the 1.6~ 
$\mu$m region. Figure \ref{fig:Devi} compares our results and HITRAN2012 line 
intensities to these new experimental results. The HITRAN2012 data comes from 
the CDSD database. A significant systematic shift of $5\%$ and $10\%$ toward 
higher intensities is observed for both 30012 -- 00001 and 30013 -- 00001 band. 
An almost identical pattern is followed by our line intensities and the 
effective Hamiltonian calculations given under HITRAN2012 entries. 
Possible problems in measured line intensities were found at $m = +38 $ and $ m = -40 $. 
These transitions clearly stand out in the comparison pattern for both HITRAN 
and the present study. High $J$ tails of both bands are bent in a bow-like 
structure, behavior which has been already observed for the main isotopologue. We 
attribute such behaviour to limited flexibility of functional form  assumed  for 
the Herman-Wallis factors, when reducing the experimental data. 

Less pronounced deviations in high $J$ region have been observed in the work by 
Benner \etal\  \cite{16BeDeSu.CO2}, where highly accurate measurements for the 
2.06~$\mu$m band of the main isotopologue were made. Comparison shows 0.5\% 
agreement between the experiment and UCL-IAO line intensities for the 20013 -- 
00001 band for the main isotopologue \cite{16BeDeSu.CO2}, and systematic 
increase in intensity deviation from $m = 60$ onwards to reach 1.5\% deviation 
at $m = 84$. Both experiments (Devi \etal\  \cite{16DeBeSu.CO2} and Benner \etal\  
\cite{16BeDeSu.CO2}) utilized the same multispectrum nonlinear least 
squares curve fitting technique to retrieve line profiles and intensities. This similarity in 
high $J$ behaviour supports the thesis of potential problems with retrieval 
model used in experimental post processing.

The accuracy of the experiments by Devi \etal\ and the present calculations has been verified by very recent Cavity Ring-Down Spectroscopy measurements of CO$_2$ lines by Kiseleva \etal\ \cite{16KiMaSt.CO2}. Their observed intensity of the P(6) line in the 30013 -- 00001 band of the 636 isotopologue was found to be within 0.4\% of both UCL and HITRAN line intensities. Further proof of the consistency between the experiment by Kiseleva \etal\ and the present study is the P(52) line of the 30014 -- 00001 band of the main isotopologue (626), which was also compared and found to be only 0.17\% from our predicted value. Both lines were measured with stated $<$1\% uncertainty budget. This suggests that a similar, presumably sub-percent, accuracy for the line intensities provided here and by HITRAN2012 for the 30013 -- 00001 band of the 636 isotopologue.

\begin{figure}[H]
  \includegraphics[width=14cm]{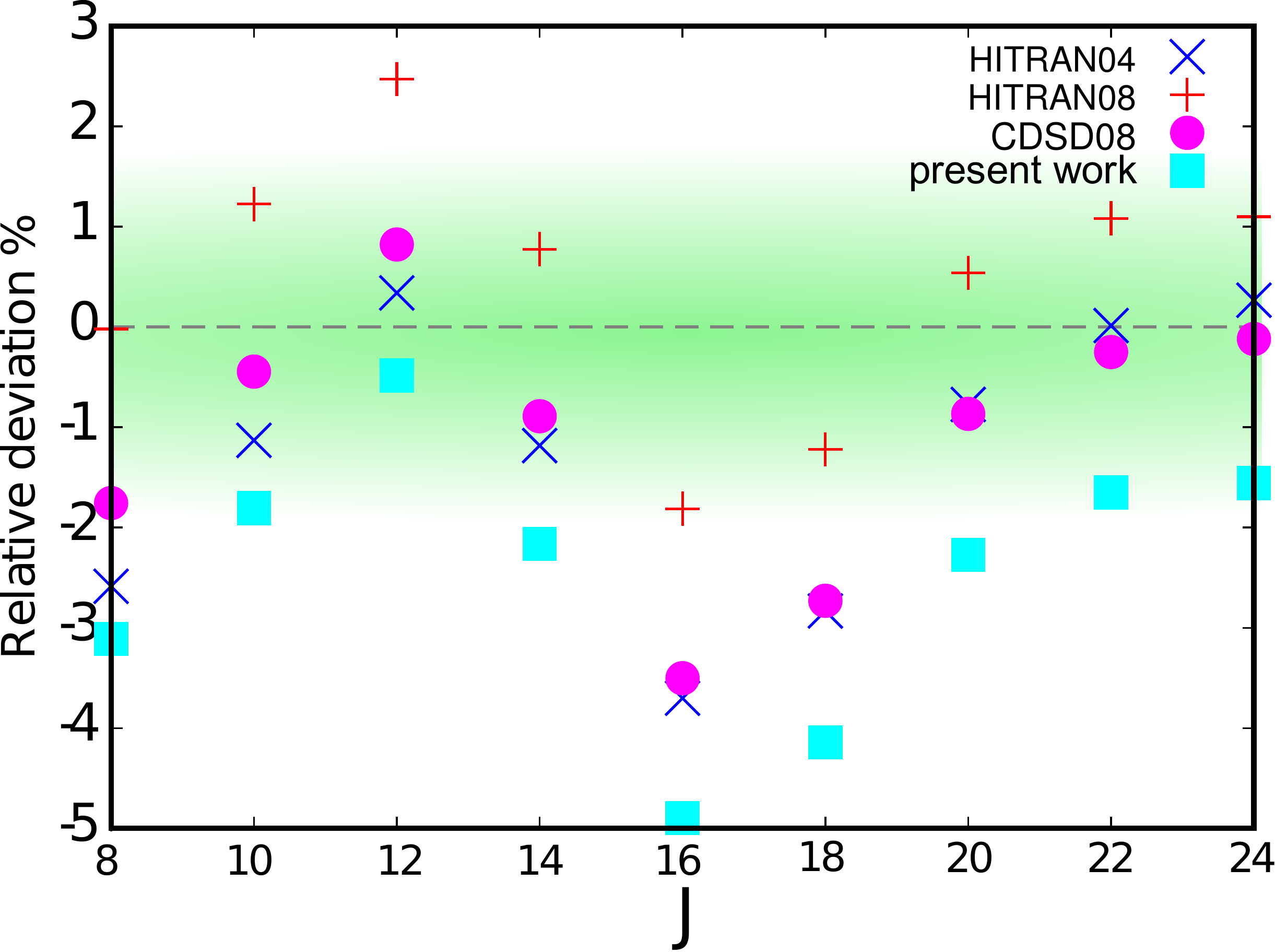}
\caption{Relative deviations of line intensities of the 20012 -- 00001 band of 
the 636 isotopologue from measurements by Durry \etal\ \cite{10DuLiVi.CO2} 
plotted against $J$ quantum number for several databases. Sources
considered are  HITRAN2004 \cite{jt350}, 
HITRAN2008 \cite{jt453}, the 2008 release of CDSD \cite{08PeTa.CO2}
and the present work.
The 1\% deviation region 
is represented by green edge-blurred strip. }
\centering
\label{fig:Durry}
\end{figure}

The 2010 experimental study by Durry \etal\ \cite{10DuLiVi.CO2} deserves 
special attention, as intensity uncertainties for measured bands of the 636 
isotopologue are claimed at the 1\% level. Figure \ref{fig:Durry} compares 
experimental line intensities from Durry \etal\ with  HITRAN2004 \cite{jt350}, 
HITRAN2008 \cite{jt453} and 2008 release of the CDSD database 
\cite{08PeTa.CO2,11TaPe.CO2}, as well as with our calculated values. A 
characteristic wave-like pattern is visible. As all four sources follow this 
envelope, but with different systematic shift, we conclude this 
pattern is an artifact of the results of Durry \etal. The points from present work are shifted toward 
most negative values of relative deviation, with average systematic shift of $~$2\%. 
Intensity comparison for this band for the main 626 isotopologue \cite{16BeDeSu.CO2} supports the 1\% accuracy our theoretical intensities. 
Therefore it would seem that the stated 1\% uncertainty of Durry \etal's measurements may be too optimistic.

\subsection{Isotopologue 727}

\begin{figure}[H]
  \includegraphics[width=14cm]{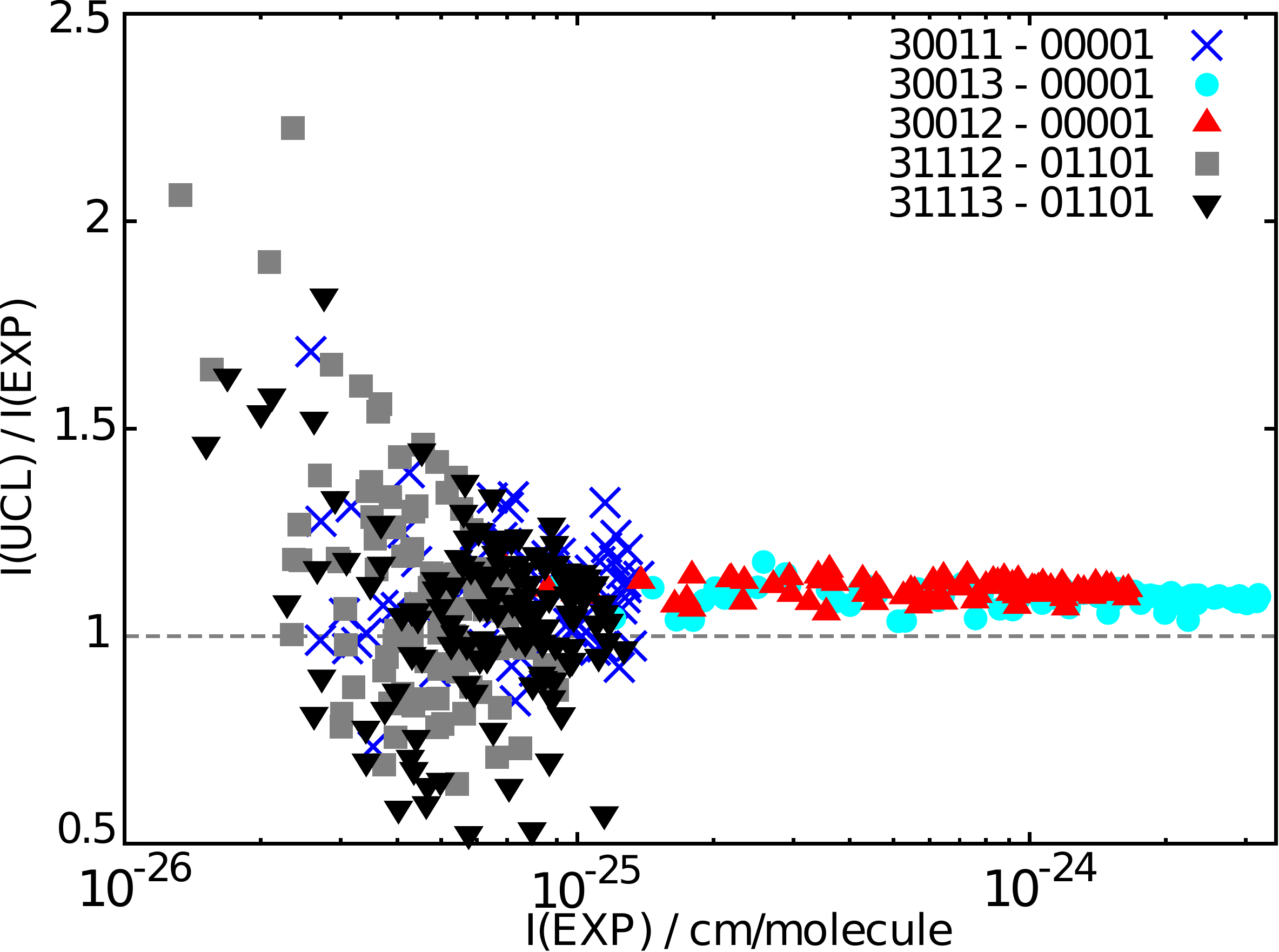}
\caption{Relative intensities for several bands of the 727 isotopologue measured by Jacquemart \etal\ \cite{15JaBoLy.CO2}.}
\centering
\label{fig:727_Jacquemart}
\end{figure}

In recent measurements performed on $^{17}$O and $^{18}$O enriched samples, 
Jacquemart \etal\  \cite{15JaBoLy.CO2} measured several bands for the 727 
isotopologue. The authors argue that only lines stronger than 10$^{-25}$ 
cm/molecule are retrieved with 'good accuracy' and this accuracy is also 
strongly dependent on the knowledge of isotopic abundances.  Figure 
\ref{fig:727_Jacquemart} compares intensities of different bands 
measured by Jacquemart \etal\ with our predictions. It is evident that lines weaker than 
$1.0\times10^{-25}$ cm/molecule give reduced accuracy, as statistical spread 
appears an order-of-magnitude larger than for the strong bands measured in this 
experiment. Hence the experimental results \cite{15JaBoLy.CO2} for the 30011 -- 00001, 31112 -- 01101 and 31113 -- 01101 band should be considered with caution.

The intensities of the two strongest bands in the 2~$\mu$m region,
that is 30012 -- 00001 and 30013 -- 00001, are uniformly shifted by
+10\% with respect to experiment.  As indicated by Jacquemart \etal\ \citep{15JaBoLy.CO2}, intensities of whole bands are strongly dependent on
isotopologue abundance (reported as 22.27\%), and this factor is
considered to be the main source of possible systematic shifts with
respect to other studies. Comparisons with previous measurements by
Karlovets \etal\ \cite{14KaCaMo.CO2} were made, revealing the new
measurements by Jacquemart \etal\ \citep{15JaBoLy.CO2} to be on
average 3 -- 4\% stronger. However, samples used by Karlovets \etal\
had very low abundance of 727 (0.04\%), which resulted in large
statistical error (15\%) in the intensities.  Therefore we conclude
that with the current level of experimental control over systematic
errors it is difficult to reliably refer to measurements better than
10\% accuracy. Nonetheless, because theoretical line intensities have
constant accuracy for whole bands (except resonances), they can be
used to assess the precision of measurements. Small scatter of line
intensities throughout these bands (marked red and cyan in Figure
\ref{fig:727_Jacquemart}) confirms the claimed high precision (1\%) of
the measurement from Ref. \citep{15JaBoLy.CO2} above
$2.5\times10^{-24}$ cm/molecule, ~2\% between $5\times10^{-25}$ and
$2.5\times10^{-24}$, ~5\% between $1\times10^{-25}$ and
$5\times10^{-25}$, and ~20\% below $1.0\times10^{-25}$.

\begin{figure}[H]
  \includegraphics[width=14cm]{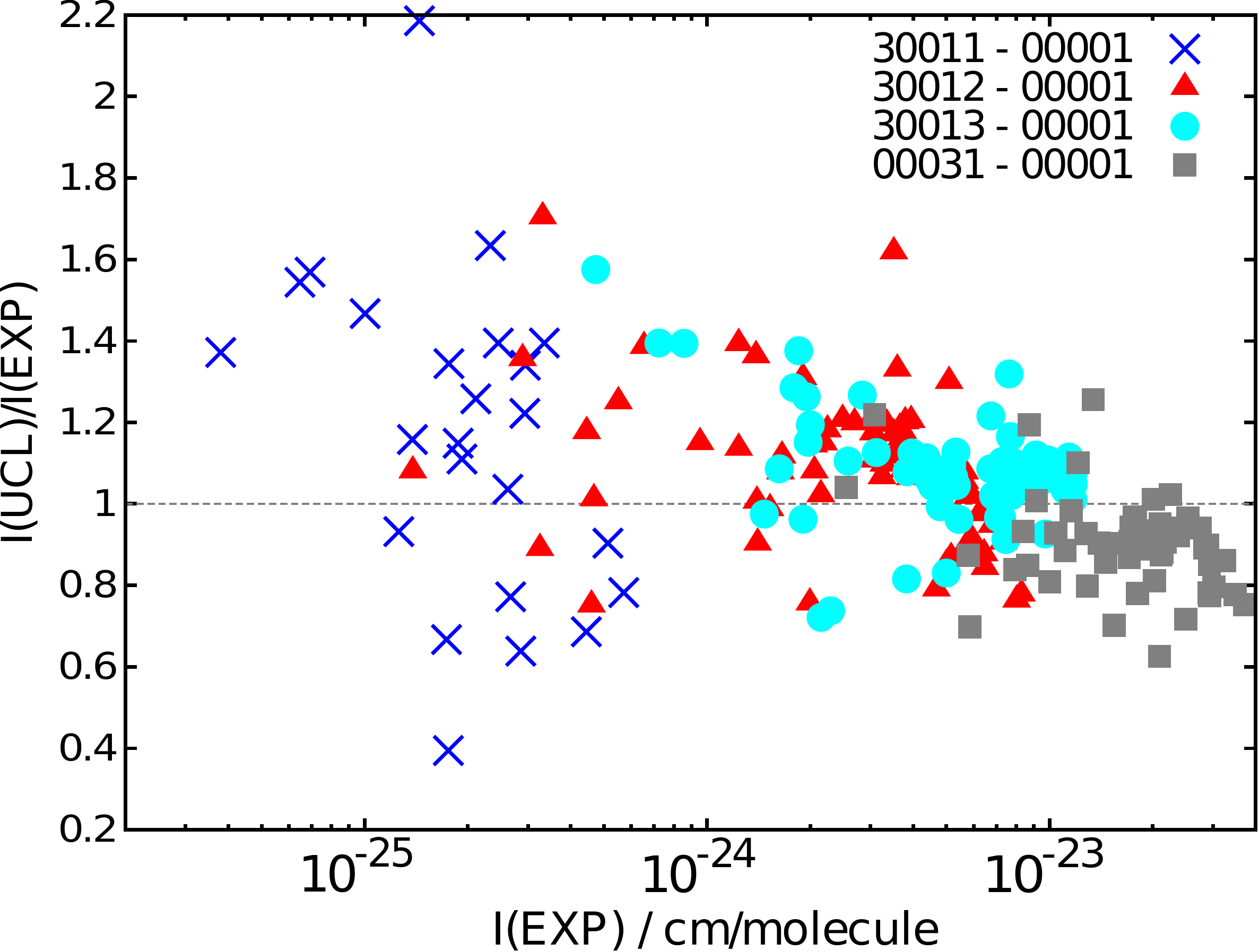}
\caption{Relative intensities for several bands of the 727 isotopologue measured by Karlovets \etal\ \cite{14KaCaMo.CO2}. Intensities were scaled to unit abundance.}
\centering
\label{fig:727_Karlovets}
\end{figure}

Figure \ref{fig:727_Karlovets} compares line intensities from the
present work to the experimental data from Karlovets \etal\
\cite{14KaCaMo.CO2}. The low isotopic abundance of samples used in
experiments and large stated uncertainty (15\%) means that the
comparison despite its large scatter is satisfactory. As for other
isotopologues, the 00031 -- 00001 band computed by us has an
underestimated intensity (grey squares in Figure
\ref{fig:727_Karlovets}).  Cyan and red points correspond to 30013 --
00001 and 30012 -- 00001 bands, and these experimental points were
used to relate the line intensities of these bands in the study by
Jacquemert \etal.

\subsection{Isotopologue 828}

\begin{figure}[H]
  \includegraphics[width=14cm]{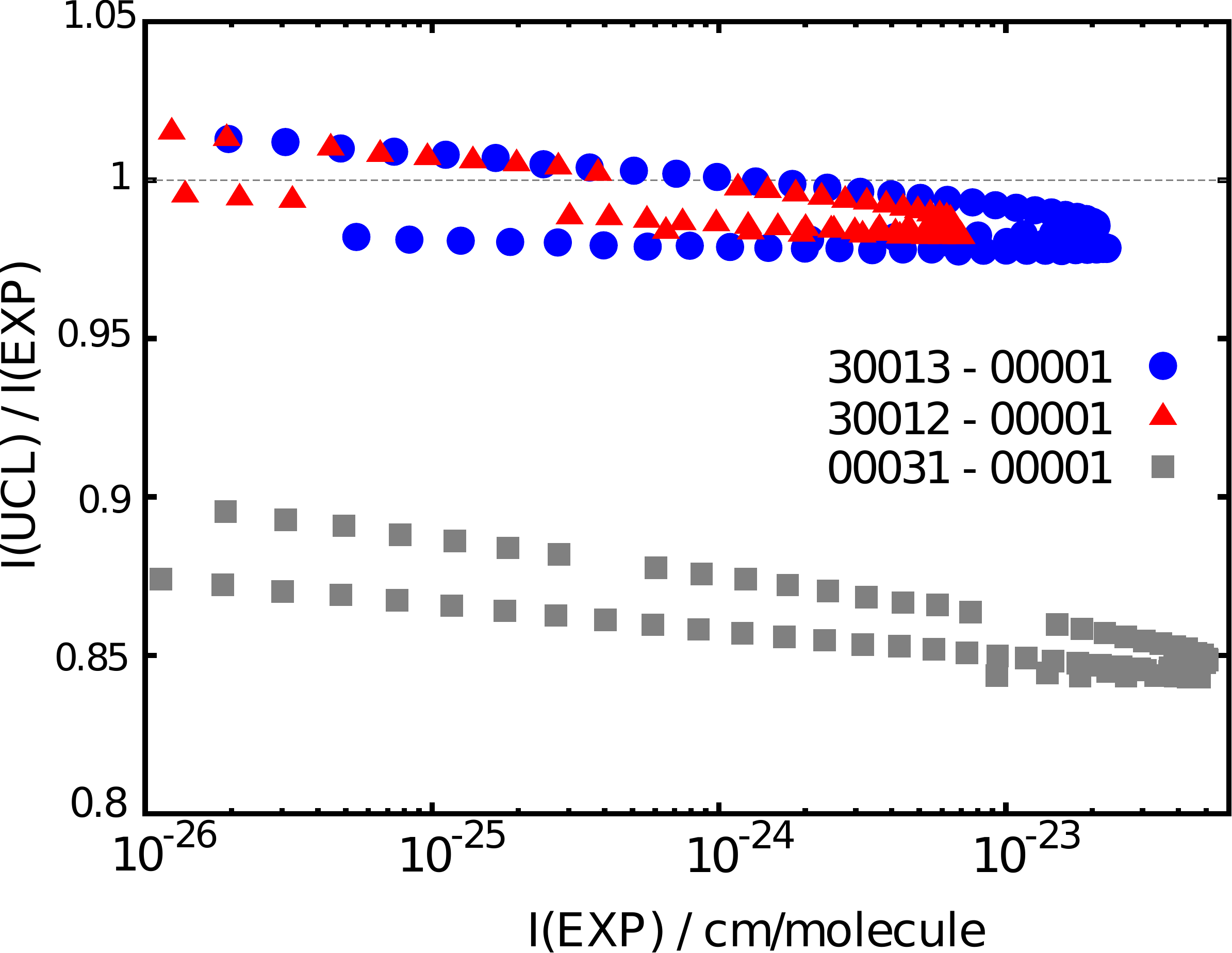}
\caption{Relative intensities for several bands of the 828 isotopologue measured by Karlovets \etal\ \cite{14KaCaMoII.CO2}. Intensities were scaled to unit abundance.}
\centering
\label{fig:828_Karlovets}
\end{figure}

Recent CW-Cavity Ring Down experiments for enriched sample of the 
$^{12}$C$^{18}$O$_2$ isotopologue by Karlovets \etal\ 
\cite{14KaCaMoII.CO2} cover the spectral range of all previous 
measurements for $\Delta P=9$ transitions. The study comprises 2870 lines from 
59 bands in the 5851 -- 6990 \cm\ region and was recorded for 25.45 \% 
abundance. Crude experimental data was fitted with an effective operator model to 
take into account another accurate experimental dataset from Toth \etal\ 
\cite{07ToMiBr.CO2}, which has been also included in the 2012 release of the 
HITRAN database. The estimated 10\% uncertainty of the line intensities is the most 
accurate claim up-to-date. For a detailed review of previous measurements for this 
isotopologue see Refs.\cite{14KaCaMoII.CO2,15TaPeGa.CO2} and references 
therein. Here, highly enriched sample allowed for more precise measurements than 
in the 727 isotopologue case. 
Figure \ref{fig:828_Karlovets} compares line intensities of the three strongest 
bands measured by Karlovets \etal\ to present study. The 30012 -- 00001 and 
30003 -- 00001 bands remain within $\pm 2\%$ deviation range,
which suggest that the stated experimental uncertainty of 10 \% is actually too 
pessimistic. Our intensities for the 00031 -- 00001 band are shifted down by 14 
\%, similar to other isotopologues. The recommended line list uses line 
intensities from CDSD-296 for this band.

Above results for 636, 727 and 828 isotopologues are summarized in Table \ref{table:exp}.

\begin{table}[ht]
\caption{Characterization of selected vibrational bands of three symmetric CO$_2$ isotopologues. 
Given for each band and each reference are the number of lines in the band,  accuracy declared in the reference, 
average systematic shift ($\Delta_{sys}=\bar{S}$: average residual with respect to present study), average statistical
dispersion ( $\Delta_{stat}=\sqrt{\frac{1}{N_{lin}}\sum_{i=1}^{N_{lin}}\left(S_i-\bar{S}\right)^2}, $\quad $ S_i=\left|\frac{I_{UCL,(i)}}{I_{exp,(i)}}-1\right|\cdot 100\%$) 
and the total band strength in cm/molecule. The last column (marked UCL-IAO) contains the data from the present study, 
the total number of lines in the band, suggested accuracy for the band (in \%) and the total band strength in cm/molecule.}

\scriptsize
\begin{tabular}{l l l l l l l l l l l }
\hline\hline
Iso. & Band & $N_{lin}$ & Strength  & acc. (\%)  &  $\Delta_{sys}$(\%) &  $\Delta_{stat}$ (\%) & & $N_{tot}$ & Strength  & acc. (\%) \\\cmidrule{3-7}\cmidrule{9-11} 

 &	&	&	\multicolumn{2}{l}{Karlovets \etal\ (2013) \cite{13KaCaMo.CO2}} 	& &	&&	\multicolumn{2}{r}{UCL-IAO}& 	\\\cmidrule{3-7}\cmidrule{9-11}

727 & 30012--00001 & 64 & $2.13\times10^{-22}$ & 3-20& +17 & 13  &   & 64 & $2.22\times10^{-22}$ & 1 \\ [0.5ex] 
	& 30013--00001 & 58 & $3.48\times10^{-22}$ & 3-20 & +13 & 12  &  &  58 & $3.71\times10^{-22}$ & 1 \\ [0.5ex] 

 &	&	&		\multicolumn{2}{l}{Jacquemart \etal\ (2015) \cite{15JaBoLy.CO2}} & &	&	&	& & \\\cmidrule{3-7}\cmidrule{9-11}

727	& 30012--00001 & 85 & $6.85\times10^{-23}$ & 20  	& +11 & 2  & & 85 & $7.59\times10^{-23}$ & 1 \\ [0.5ex] 
	& 30013--00001 & 93 & $1.37\times10^{-22}$ & 20 		&  +9 & 2  & & 93 & $1.50\times10^{-22}$ & 1 \\ [0.5ex] 
    & 31113--01101 & 130 & $8.84\times10^{-24}$ & $>$20 	& +17 & 15 & & 130 & $9.21\times10^{-24}$ & 3 \\ [0.5ex] 

   &	&	&		\multicolumn{2}{l}{Karlovets \etal\ (2013) \cite{13KaCaMo.CO2}} &	& &	&	& &  \\\cmidrule{3-7}\cmidrule{9-11}%

828 & 30012--00001 & 64 & $1.86\times10^{-22}$ & 10 & -2  & 2  &   &	64 & $1.83\times10^{-22}$ & 1 \\ [0.5ex] 
	& 30013--00001 & 81 & $6.05\times10^{-22}$ & 10 &  -2 & 3  &   &	81 & $6.16\times10^{-22}$ & 1 \\ [0.5ex] 
    & 00031--00001 & 80 & $1.33\times10^{-21}$ & 10 & -13 &  5 &   & 80 & $1.13\times10^{-21}$ & 20  \\ [0.5ex] 
    
 &	&	&		\multicolumn{2}{l}{Devi \etal\ (2016) \cite{16DeBeSu.CO2}} &	&&	&	& & 	 \\\cmidrule{3-7}\cmidrule{9-11}%
    
636 & 30012--00001 & 55 & $5.41\times10^{-24}$ & 10 & +4 & 3   &  &  55 & $5.67\times10^{-24}$ & 1 \\ [0.5ex] 
	& 30013--00001 & 47 & $2.03\times10^{-24}$ & 10 & +8 & 15  &  &  47 & $2.18\times10^{-24}$ & 1 \\ [0.5ex]
\hline\hline
\end{tabular}
\label{table:exp}
\end{table}

\subsection{Comparison with HITRAN2012, Ames and CDSD-296}

The HITRAN2012 database comprise line lists for 636, 727, 828 and 838. Uncertainty 
indices of the line positions range from 2 ( $\geq$ 0.01 \cm\ and $<$ 0.1 \cm) to 
9 ($\geq$   $10^{-9}$ \cm and $<$ $10^{-8}$ \cm).  In general, line positions 
from the latest version of CDSD-296 are very close to the line positions given in 
HITRAN2012 and have uncertainties corresponding to indices ranging from 3 to 9 
depending on spectral region and quality of underlying experimental entries. 
Intensities provided by the current release of HITRAN for symmetric isotopologues of 
carbon dioxide come from two main sources: experiment (JPL OCO line list) by 
Toth \etal\ \cite{08ToBrMi.CO2} and the majority of transitions from a previous 
version of CDSD.  The estimated uncertainties for all CDSD intensities is given 
as 20 \%\ or worse in HITRAN (uncertainty code 3). However, this number does not 
reflect the actual uncertainties of the intensities. Most of the HITRAN 
intensities have the uncertainties much better than 20\%. More detailed 
information about the actual uncertainties can be found in the official release 
of CDSD \cite{15TaPeGa.CO2}, which can be used
to get more
realistic information about the uncertainties of the line parameters.

Intensities from Toth \etal are supposed to be accurate to better than 2\%\ 
(uncertainty code 7) or 5\%\ (code 6). 
As discussed elsewhere \cite{15ZaTePo.CO2}, the stated uncertainty estimates of 
all current entries are insufficiently  accurate for remote sensing 
applications. In addition to that, several bands feature unrealistic jumps in 
line intensities originating from switching data sources.  
Therefore a unified 
approach giving line positions of spectroscopic quality combined with 
significantly more accurate transition intensities is needed. 
In the present paper we aim in fulfilling these requirements.

In order to relate results from the present study to data given in HITRAN we 
compared line intensities for matched lines between the two line lists (see Table 
\ref{table:info}). As a primary measure of relative intensity deviation from 
HITRAN data we used the following formula:
\begin{equation}
S=\left(\frac{I_{UCL}}{I_{HIT}}-1\right)\cdot 100 \%
\end{equation}
where $I_{UCL}$ stands for line intensity from UCL-IAO line list given in 
cm/molecule and $I_{HIT}$ is HITRAN2012 intensity.

This measure is adequate for small deviations but poorly illustrates 
highly discrepant intensities, due to its asymmetric functional form. For larger 
deviations, especially for line intensities weaker than HITRAN by more than 
100\%, the quasi-symmetry is noticeably broken, resulting in a biased picture. 
In such cases, for example to show graphically a general overview, we decided to 
use a symmetrized measure to account for proper representation of large 
deviations:

\begin{equation}
S_{sym}=\left(\frac{I_{UCL}}{I_{HIT}}-\frac{I_{HIT}}{I_{UCL}}\right)\cdot 100 \%
\label{eq:symmetric}
\end{equation} 
This measure, in turn, yields far from intuitive numbers near $0\%$ deviation.

\subsubsection{Isotopologue 636}
The HITRAN2012 line list for the second most abundant 636 isotopologue contains 
68~856 lines below 8000 \cm. There are two sources of line intensities: the 
majority of lines taken from  the 2008 version of the CDSD-296 database 
\cite{08PeTa.CO2} and two bands (20012--00001 and 20013--00001) from high 
precision measurements by Toth \etal\ \cite{08ToBrMi.CO2}. All lines present in 
the HITRAN2012 database for this isotopologue were matched to our line list with 
a root mean squared deviation (RMSD) of 0.04 \cm. Lines that lay far in intensity from our predictions 
(\textit{vide infra}) were double checked by manual investigation.

\begin{figure}[H]
    \includegraphics[width=\textwidth]{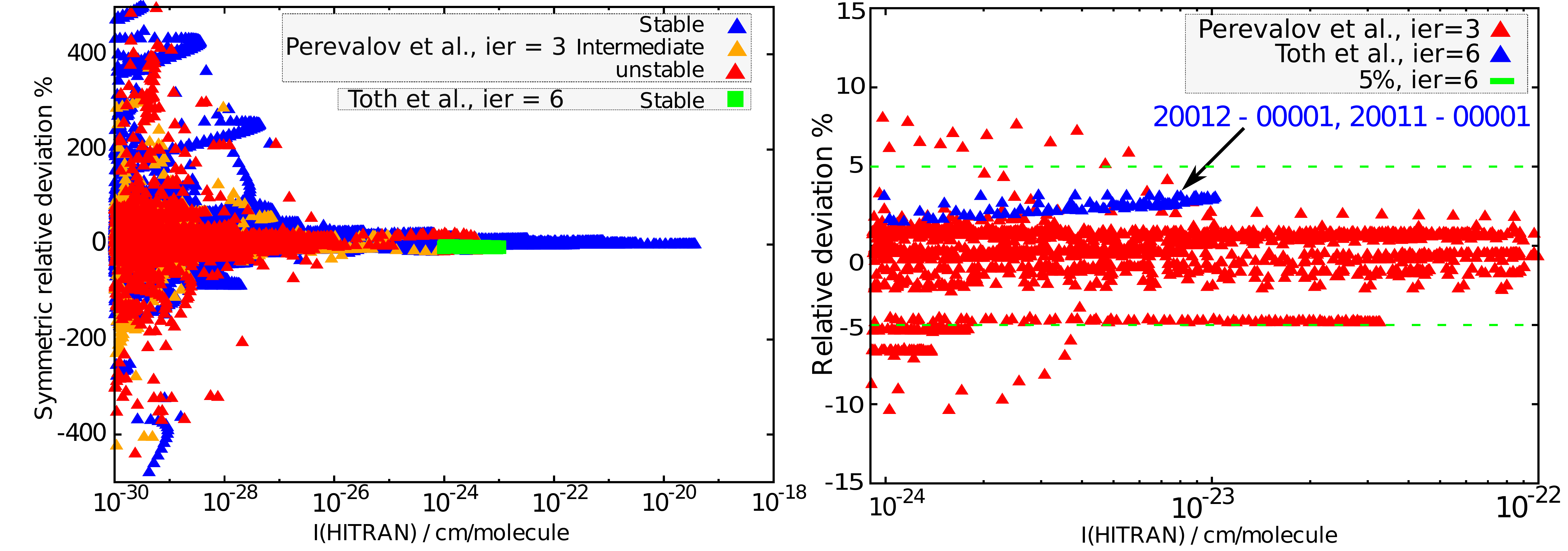}
    \caption{Left panel represents symmetric relative deviation for the 636 isotopologue for the two different sources (Perevalov \etal\ \cite{08PeTa.CO2} and Toth \etal\ \cite{08ToBrMi.CO2}) from the HITRAN2012 database. Right panel is a zoomed image in the region of high accuracy ($ier = 6$) measurement by Toth \etal\ . Dashed green line indicates 5\% limit of deviation tolerance associated with $ier = 6$. Two bands measured by Toth \etal\ are marked with arrow }
  \label{fig:636}
\end{figure}

An overview from Figure~\ref{fig:636} reveals the rather typical situation of 
funnel shaped relative deviation plot. By zooming into the region of high 
accuracy measurement by Toth \etal, one can clearly see that all lines (marked 
with blue filled triangles in the right panel of Figure \ref{fig:636}) remain 
within the claimed 5\% uncertainty, additionally exhibiting a very narrow spread.

\subsubsection{Isotopologue 727} 
\begin{figure}[H]
  \includegraphics[width=14cm]{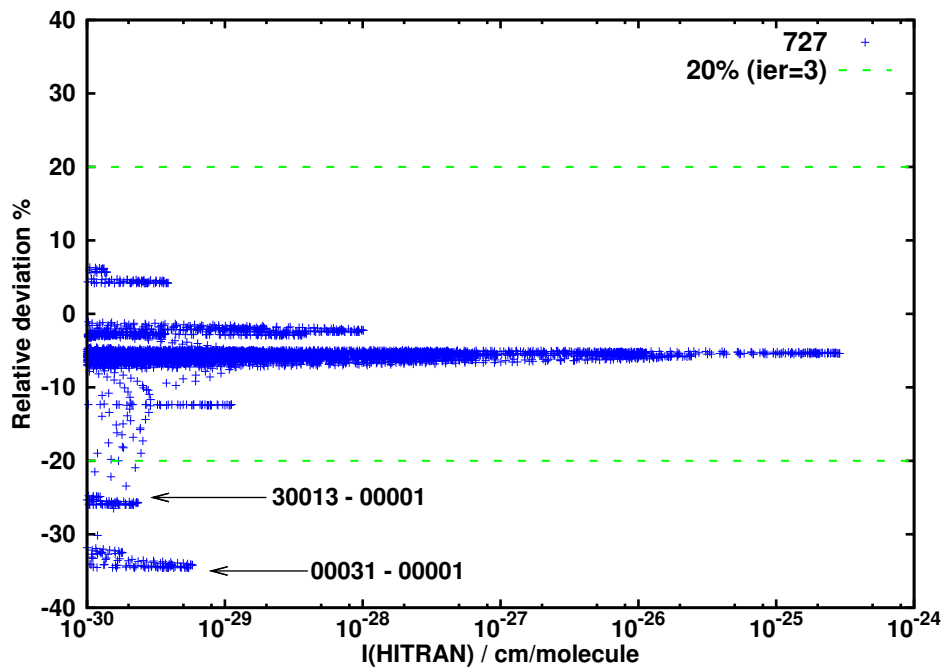}
\caption{Relative intensities  (cf. eq. (\ref{eq:symmetric})) plotted against HITRAN2012 line intensities for the 727 isotopologue. Green dashed horizontal line represents deviation from HITRAN2012 data equal to $\pm 20$\%.}
\centering
\label{fig:727}
\end{figure}

HITRAN2012 line list for the 727 isotopologue contains 5187 lines below 8000 
\cm, all of which were taken from the effective Hamiltonian calculations by 
Tashkun and Perevalov \cite{12TaPe_dip.CO2}. Figure \ref{fig:727}
compares our line intensities (all stable) to HITRAN2012; we observe the majority of line intensities 
display a systematic shift of -6\% with respect to those recommended by 
HITRAN. Here again, noticeable arc structures appear. Similar 
behaviour was observed for the main 626 isotopologue. Although most of arcs are 
rather flat, there are a few bands which arc structure extends over a wide 
deviation range. Such occurrences may be caused by insufficiently flexible functional form of the Herman-Wallis 
factors employed to reduce experimental data for those bands, resulting in inaccurately retrieved experimental line 
intensities, especially for high $J$s. These serve as an input to the effective Hamiltonian calculations (CDSD, hence HITRAN),
thus artifacts of experimental analysis are likely to be propagated within the EH approach. 

We have already shown that inaccuracies of our model are largely reflected in 
systematic shifts of whole bands, rather than statistical scatter, which is 
assumed to remain almost constant as a function of $J$. Two bands in Figure
\ref{fig:727} lie outside the tolerance given by the HITRAN2012 uncertainty 
code 3. These are: the 00031 -- 00001 band and the 30013 -- 00001 band (both 
indicated with arrows in Figure \ref{fig:727}). The discrepancy for the former band 
has been explained in terms of rather poor reproduction of the 3$\nu_3$ series 
of bands by our DMS and needs to be replaced in our recommended line list. The 
behavior of the latter band however is not clearly understood at this stage and 
requires further investigation. Our working hypothesis is that the $-$6\% 
systematic shift applies to all bands, hence the 30013 -- 00001 band when 
shifted by +6\%, should match the 20\% tolerance 
region, which is also regarded as provisional. New measurements by Karlovets \etal\ \cite{14KaCaMo.CO2} 
and Jacquemart \etal\ \cite{15JaBoLy.CO2} analysed in  the previous section improved the quality of the line positions 
and intensities for the 30013 -- 00001 band.

\subsubsection{Isotopologue 828}
\begin{figure}[H]
  \includegraphics[width=\textwidth]{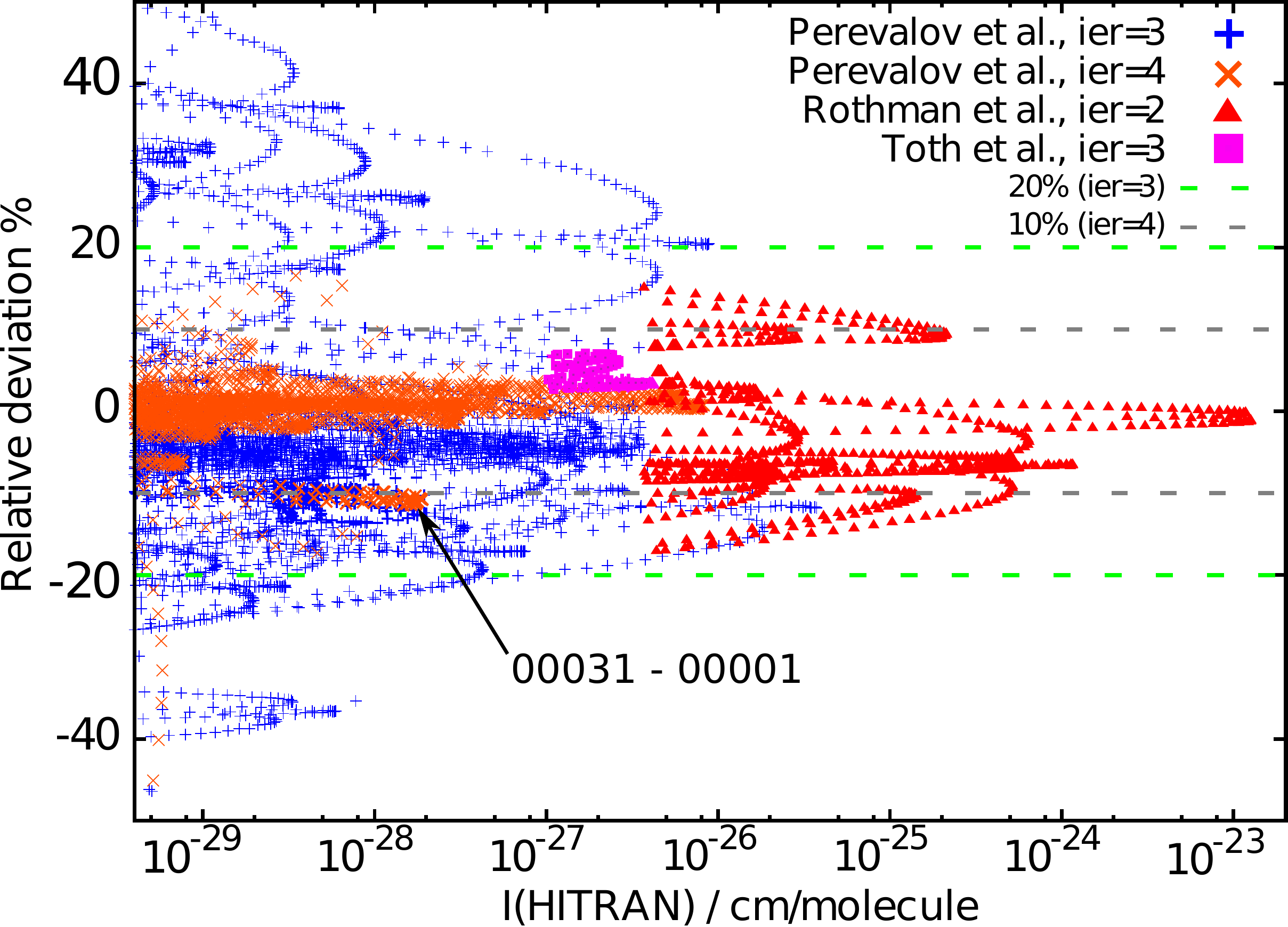}
\caption{Relative intensities from the present study plotted against HITRAN2012 line intensities for the 828 isotopologue. Only $\pm 50\%$ region is depicted. Dashed grey and green lines correspond to 10\%  and 20\% deviation, respectively. Blue crosses correspond to a subset of lines taken from Perevalov \etal\ \cite{08PeTa.CO2} which has been assigned to $ier = 4$. Consequently, rotated orange crosses represent $ier = 3$ from the same reference. Red filled triangles refer to Rothman \etal\  \cite{92RoHaWa.CO2}, while purple filled squares stand for the small set of lines provided by Toth \etal\ \citep{08ToBrMi.CO2}.}
\centering
\label{fig:828_sources}
\end{figure}

HITRAN2012 line list for the 828 isotopologue contains 7071 lines below 8000 
\cm. There are three sources of line intensities: 6280 lines taken from 
CDSD-296 \cite{08PeTa.CO2} with $ier$ equal to 3 and 4, 722 lines taken from a
1994 update to older variational calculations \cite{92RoHaWa.CO2} with $ier$ equal to 2, and finally 69 
lines taken from measurements by Toth \etal\ \cite{08ToBrMi.CO2} with ier 
assigned to 3. Figure \ref{fig:828_sources} compares intensities from the 
present study to HITRAN2012 data. Despite the low uncertainty index, line 
intensities originating from Rothman \etal\ \cite{92RoHaWa.CO2} agree within 
$\pm 20\%$ with our results. Transitions around 2.06~$\mu$m measured by Toth 
\etal\ \cite{08ToBrMi.CO2} are enclosed in 10\% region reflecting the \textit{ier} value for this 
set. Data points originating from CDSD-296 are divided into two sets with 
differing uncertainty index. The more accurate subset (marked with orange 
rotated crosses) is clearly squeezed along the relative deviation axis and 
exhibits almost no systematic shift. In contrast, the lower accuracy subset 
from CDSD spreads over a large region in relative deviation space. This 
suggests that both sets were calculated with separate input parameters of 
different quality. The 30013 -- 00001 band ($ier = 4$) deviates around +2\% from 
CDSD predictions, while the relatively strong 00031 -- 00001 band ($ier = 4$) 
lies 11\% below the zero deviation line (visible in Figure 
\ref{fig:828_sources}). It should be noted that large deviations of the lower
accuracy CDSD-296 data ($ier = 3$) occur for very weak lines, for each the 
respective experimental data to fit the effective dipole moment parameters are 
absent. In these cases the parameters of the principal isotopologue were used in CDSD-296.

\begin{figure}[H]
  \includegraphics[width=14cm]{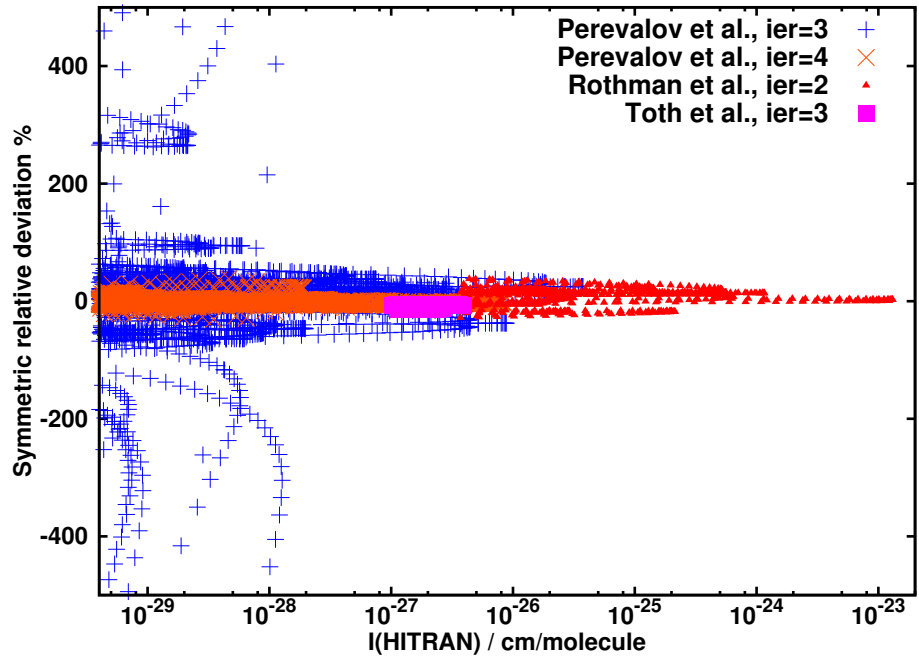}
\caption{Symmetric relative intensities (cf. eq. (\ref{eq:symmetric})) plotted against HITRAN2012 line intensities for the 828 isotopologue. Only $\pm 500\%$ region is depicted. Blue crosses correspond to a subset of lines taken from Perevalov \etal\ \cite{08PeTa.CO2} which has been assigned to $ier = 4$. Consequently, rotated orange crosses represent $ier = 3$ from the same reference. Red filled triangles refer to Rothman \etal\  \cite{92RoHaWa.CO2}, while purple filled squares stand for the small set of lines provided by Toth \etal\ \citep{08ToBrMi.CO2}}
\centering
\label{fig:828_sources2}
\end{figure}

\begin{figure}[H]
  \includegraphics[width=14cm]{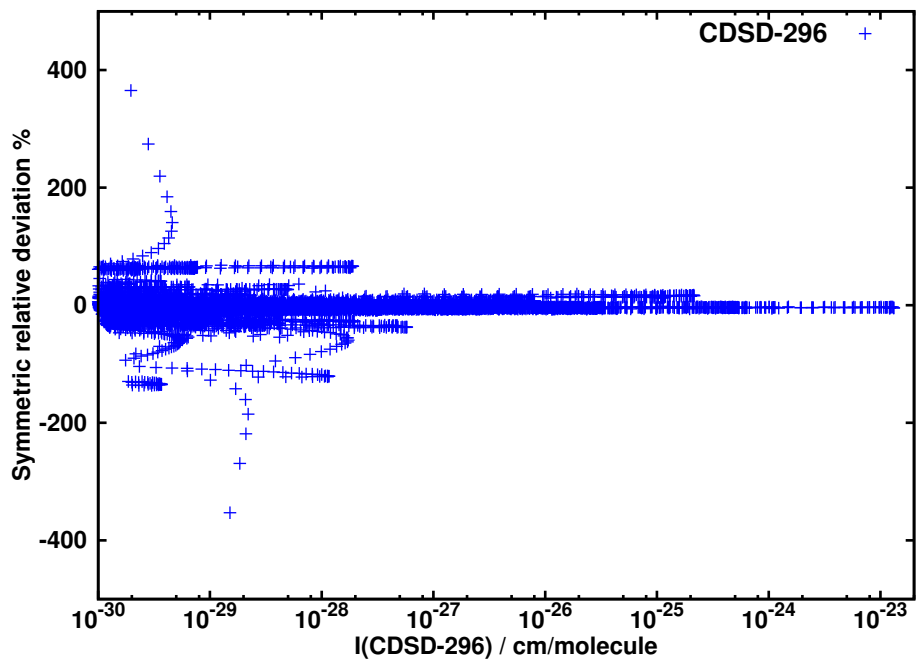}
\caption{Symmetric relative intensities  (cf. eq. (\ref{eq:symmetric})) plotted against CDSD-296 line intensities for the 828 isotopologue. Only $the \pm 500\%$ region is depicted.}
\centering
\label{fig:CDSD-UCL-828}
\end{figure}

Figure \ref{fig:828_sources2} shows that all strong lines ($>10^{-28}$ 
cm/molecule) follow funnel shape envelope, thereby reflecting the typical relation 
between intensity and accuracy of lines. However several weaker lines, which 
constitute whole bands, align in wide arc structures with large systematic 
shift. This is particularly visible for lowered accuracy lines from HITRAN2012 
(blue crosses in Figure \ref{fig:828_sources2}). These lines were directly 
incorporated from HITRAN2008. The current release of the CDSD database improved on 
accuracy of these weak lines. A comparison between UCL  and CDSD-296 intensities is 
given in Figure. \ref{fig:CDSD-UCL-828}.

\subsubsection{Isotopologue 838}
\begin{figure}[H]
  \includegraphics[width=14cm]{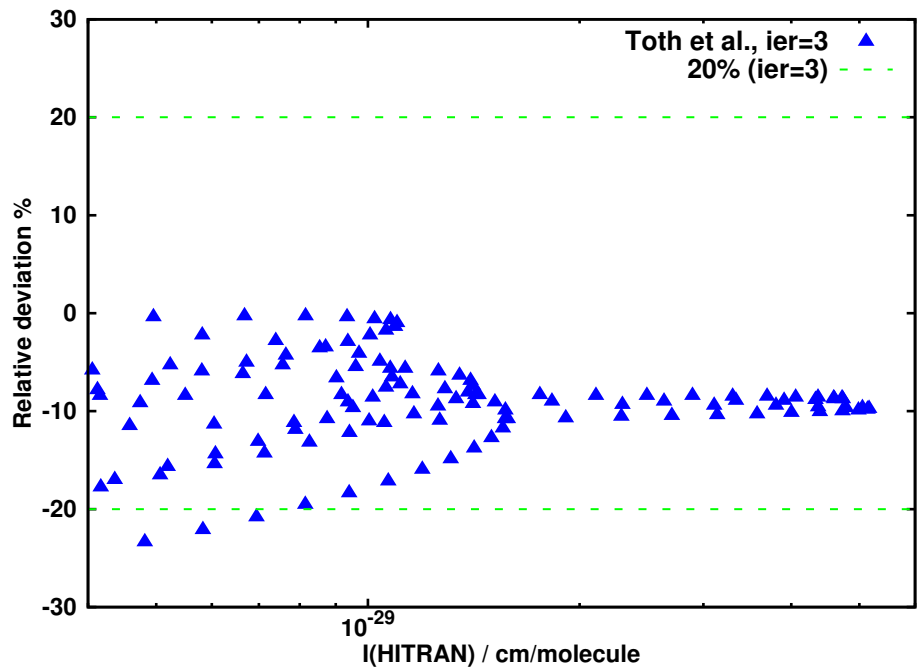}
\caption{Relative intensities plotted against HITRAN2012 line intensities for the 838 isotopologue. The triangles represent the 121 lines measured by Toth \etal\ \citep{08ToBrMi.CO2} and included in the current version of HITRAN.}
\centering
\label{fig:838}
\end{figure}

Only limited data are available for the 838 isotopologue in the 2012 edition 
of HITRAN. 121 lines measured by Toth \etal\ \cite{08ToBrMi.CO2} have 
uncertainty flag 3 and cover three bands in the 2~$\mu$m region: 20011 -- 00001, 
20012 -- 00001 and 20013 -- 0000. All lines present in this set are matched to 
our line list with a RMSD $ = 0.04$ \cm. Here, similar to the 727 case, a 
systematic shift of around 10\% is visible. This causes three transitions to 
breach the stipulated accuracy tolerance. Nevertheless, this should be 
considered as rather illusory due to the systematic shift of lines coming from 
all three bands. Figure \ref{fig:838} shows calculated intensities relative to 
HITRAN2012, where all computed lines are classified as stable.

\subsubsection{Overview}

A natural question to ask is if accuracy of our computational scheme based on 
\abinitio\ DMS holds at the same level for all symmetric isotopologues. The
current status of experimental accuracy does not give a definite answer 
to this question and there is a clear need for high accuracy measurements.
In the meantime we formulate a weaker query: do line intensities for a chosen 
band maintain similar relative deviation from  consistent, highly accurate 
experimental data source for all symmetric isotopologues? Unfortunately, 
HITRAN2012 contains no such band. Arguably the most important band from an
astrophysical and atmospheric measurements perspective is the 20012 -- 00001 
band, located in the 2~$\mu$m wavelength region. A comparative study for this 
band for different CO$_2$ symmetric isotopologues is therefore given below in 
Figure \ref{fig:20012}.

\begin{figure}[H]
  \includegraphics[width=14cm]{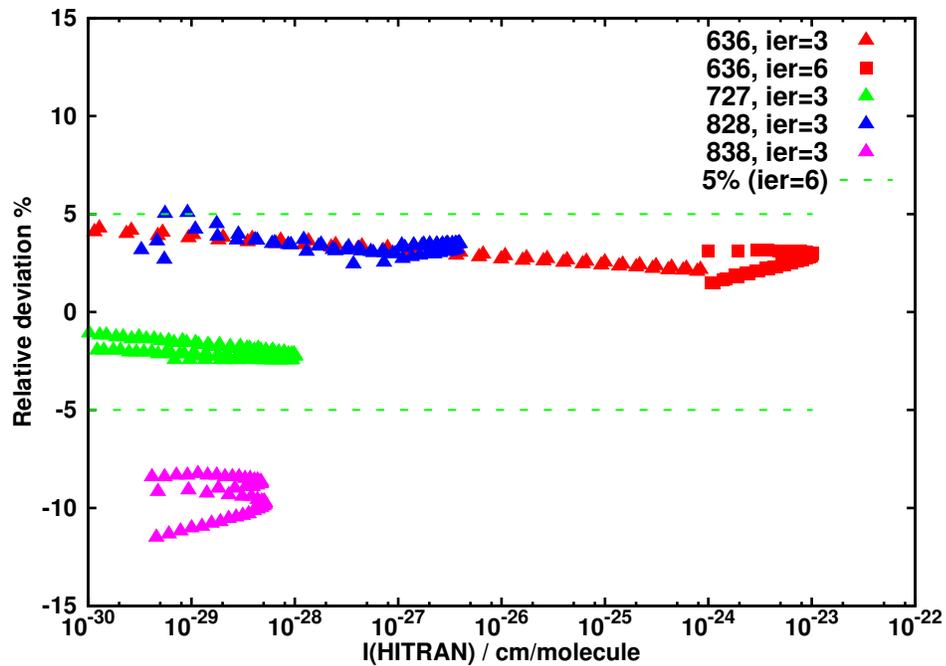}
\caption{Relative intensities plotted against HITRAN2012 line intensities for the 20012 -- 00001 band for four symmetric isotopologues. 
Red filled squares represent lines (636) measured by Toth \etal \cite{08ToBrMi.CO2} appearing with $ier = 6$. The remaining lines 
have code $ier = 3$.}
\centering
\label{fig:20012}
\end{figure}

It is instructive to follow changes in relative intensity deviation among 
different isotopologues for the selected band. 
Two data sources were used for this band: a small set of low $J$ lines from 
experiment by Toth \etal\ \cite{08ToBrMi.CO2} for the 636 isotopologue (given uncertainty index 6) 
and CDSD-296 for rest of the lines ($ier = 3$). All lines compared above match the 
stipulated HITRAN uncertainty, that is lines with $ier = 6$ fit the 5\% 
tolerance,  and the rest of the lines are 20\% or less away from HITRAN2012 
values. Minor discontinuity related to change of source of data is seen for the 
636 isotopologue. We conclude that the results illustrated in Figure \ref{fig:20012} 
give a tentative validation of the consistency of our approach with the analysed 
database.
Relatively good overall agreement between our line list and HITRAN2012, 
revealing only sporadic deviations that exceed the claimed HITRAN accuracy, but yet 
justified and facilitated with comparisons with recent and highly accurate 
measurements, allow us to draw a conclusion that replacing current HITRAN line 
intensities with our computed values would significantly increase the accuracy, 
reliability and consistency of the database.

\subsubsection{Ames-1}

In table \ref{table:Ames} we present comparison of root-mean-square-residuals of 
intensities between Ames-1 DMS, UCL DMS and CDSD-296 for 14 strongest bands of 
CO$_2$ for six symmetric isotopologues. By looking at a given isotopologue, 
a general trend is that the perpendicular bands ($\Delta l=+1,+2,...$) are in worse 
agreement with CDSD than the parallel bands.
On average, UCL-IAO based band intensities are in better agreement with CDSD 
than Ames for 626 and 636. the 828 isotopologue exhibits a similar level of agreement, 
and larger deviations for UCL-IAO bands are observed for 727, 838 and 737. This 
may have two causes: a) The UCL treatment does not include any beyond Born-Oppenheimer 
correction, which may be a source of errors for other than main isotopologue, 
against which the DMS was benchmarked; b) CDSD entries, as experimentally tuned, 
are less accurate for less abundant isotopologues, and derived uncertainties do 
not allow us to judge in favour of either DMS. On balance we believe the second
explanation is more likely.

\begin{landscape}
\begin{table}[H]
\caption{DMS statistics for 14 strongest carbon dioxide bands for six symmetric isotopologues. Numbers in columns correspond to root-mean-square-deviations of band intensities from the CDSD-296 database.}
\flushleft
\scriptsize
\begin{tabular}{l l l l l l l l l l l l l l}
\hline\hline 

\multicolumn{2}{l}{Isotopologue}  &\multicolumn{2}{c}{626} &	\multicolumn{2}{c}{636} & \multicolumn{2}{c}{828}  & \multicolumn{2}{c}{727}  & \multicolumn{2}{c}{838}  & \multicolumn{2}{c}{737}\\[0.5ex] 

Band & Stability &  UCL-IAO &  Ames & UCL-IAO &  Ames & UCL-IAO &  Ames & UCL-IAO &  Ames & UCL-IAO &  Ames & UCL-IAO &  Ames  \\ [0.5ex] 
\hline \\

00011 -- 00001 & Stable & 0.2 & 2.1  & 0.8 & 2.5  & 1.9 &  0.5  & 4.6 & 2.3 & 5.7  & 3.4 & 4.9 & 1.8\\ [0.5ex] 
01101 -- 00001 & Stable & 1.9 & 1.7  & 2.9 & 2.6 & 2.3 & 2.4 & 5.9   & 1.4  & 4.9   &  4.7&  5.0& 4.0\\ [0.5ex] 
01111 -- 01101 & Stable & 0.2 & 2.2	& 0.7 &	2.5 & 1.8 & 0.8 & 4.6 & 2.2 & 5.6 & 3.7 & 4.9 & 1.7\\ [0.5ex] 
02201 -- 01101 & Stable & 2.2  &  2.1 	& 7.3 &	7.0 & 2.7 & 2.5 & 6.3  &  1.6 & 5.4  & 5.0 & 5.3  & 4.2\\ [0.5ex] 
02211 -- 02201 & Stable &  0.2  &  2.2&  0.7 & 2.6 & 1.8  &  0.6  &  4.5 & 2.3  &  5.7  & 3.3 & 4.9 & 1.7  \\ [0.5ex] 
03301 -- 01101 & Stable &  2.5  &  2.2  	&  11.2 &	10.9 &   3.1 &   2.9  & 6.6  & 2.0   & 5.8 & 5.5	 & - & - \\ [0.5ex] 
10001 -- 01101 & Stable &  2.0   &  1.7  	& 0.5  &0.9 	 & 2.5 &  2.2 & 6.1   &  1.2 & 5.0  &4.3  &  4.9 & 3.4 \\ [0.5ex] 
10002 -- 01101 & Stable &  1.7 &  1.9   	& 1.9 &	 2.1 & 2.4 &  2.5 &  5.9   & 1.5   &   5.3 & 5.3  & 5.2 &  4.3 \\ [0.5ex] 
10011 -- 00001 & Stable &  0.7  &   0.8 	& 4.3  & 4.1	 & 6.9 &  7.1&  4.9  & 9.4   &  3.1 & 3.0 &  4.7 &  3.7\\ [0.5ex] 
10011 -- 10001 & Stable &   0.3 &    2.1	&  0.7  &	 2.7  & 1.9 & 0.5  & 4.6  &  2.2 & 5.7 & 3.3  & 4.9 & 1.7 \\ [0.5ex] 
10012 -- 00001 & Stable &  1.6  &   1.9  	& 1.6   &1.6	 & 8.3  &  9.1 &  5.4   & 10.4  &  3.5 &  2.8 & 6.4  & 4.8 \\ [0.5ex] 
10012 -- 10002 & Stable & 0.3   &   2.1 	& 0.7   &	2.6 &  1.9 & 0.5  &  4.6  &  2.2   & 5.7  &  3.3 & 4.8 &  1.7\\ [0.5ex] 
11111 -- 01101 & Stable & 0.9  &  0.9   	&4.4   &	 4.3  &  3.6  & 3.6  & 5.0  &   9.4  & 2.3 & 2.3  &  3.8  &  2.9 \\ [0.5ex] 
11112 -- 01101 & Stable &  2.1  & 2.5  	& 2.4  &	 3.0  &  3.6&   4.5  &  5.2  &  10.3   &  3.2   & 2.4    & 6.3 & 4.6  \\ [0.5ex] 
\hline\hline
\end{tabular}
\label{table:Ames}
\end{table}
\end{landscape}

\subsection{Radioactive isotopologue}

Due to its trace atmospheric abundance, $1.234(14)\times 10^{-12}$  
\cite{10LeNaKr.CO2}, only the strongest ro-vibrational absorption lines of 
radiocarbon dioxide ($^{14}$C$^{16}$O$_2$, 646) are accessible to accurate 
measurements. There are only 36 lines (all belonging to the 00011--00001 band) 
which have intensities above 10$^{-30}$ cm/molecule at room temperature, of 
which only $P(20)$ line (2209.10 \cm ) is located in a spectral region free of 
major interferences from other abundant atmospheric species like H$_2$O or CH$_4$.  For this 
reason, this line is most commonly chosen as a reference for determination of 
radioactive carbon concentrations. Therefore, the $P(20)$ line of the asymmetric 
stretching fundamental plays a distinct role in monitoring of carbon dioxide 
emission caused by fossil fuel combustion.

Although knowledge of the absolute value of the line strength to obtain the $^{14}$C 
concentrations in SCAR measurements performed on fossil samples 
\cite{13GaBaCa.CO2} can be avoided by using a more convoluted experimental 
procedure, encouraged by a successful retrieval of natural abundance of 646 by 
utilizing theoretically calculated line intensity (claimed 5\% accurate) by 
Galli \etal\ \cite{11GaBaBo.CO2}, we hope that our updated and plausibly 
sub-percent accurate intensities will be utilized in future experiments as a 
reference or calibration data. Usually, a sample to be analysed is cooled down 
to 195$~K$ or 170$~K$ in order to diminish interference effects from the nearby 
(separated by 230 MHz) line of the 636 isotopologue ($P(19)$ line of the 05511 
-- 05501 band). For this reason attention is also paid to line intensities at 
low temperature for this particular transition.
Table~\ref{table:P20}  compares intensities of the $P(20)$ and $P(40)$  lines obtained from several 
measurements and theoretical calculations together with their respective 
uncertainties.

\begin{table}[H]
\caption{Intensities of the $P(20)$ and $P(40)$ lines of the 00011 -- 00001 band for $^{14}$CO$_2$ taken from different experimental sources. }
\begin{tabular}{l c c}
\hline\hline
Reference & Temperature $K$ & Strength(uncertainty) $\times10^{-18}$ cm/molecule  \\ [0.5ex] 
\hline 
$P(20)$ Galli \etal\ \cite{11GaBaBo.CO2} & 195 & $3.10(15)$     \\
$P(20)$ Present study  & 195 & $3.07(3)$   \\
$P(20)$ Genoud \etal\ \cite{15GeVaPh.CO2} & 295 & $2.52(26)$  \\
$P(20)$ Present study & 295 & $2.82(3)$ \\
$P(20)$ Present study & 170 & $2.97(3)$	 \\
$P(40)$ McCartt \etal\ \cite{15McOgBe.CO2} & 300 & $0.627(30)$ \\
$P(40)$ Present study & 300 & $0.572(6)$ \\
\hline\hline
\end{tabular}
\label{table:P20}
\end{table}

Both $P(20)$ and $P(40)$ lines are considered stable according to our 
sensitivity analysis ($\rho=1.026$). Line intensity given by Galli \etal\ agree 
to 1\% with our value for $T = $ 195$~K$. Genoud \etal\ gives room temperature 
line intensity flagged with 10\% uncertainty, which lies 11\% below our 
prediction. The $P(20)$ line is 5 times stronger than $P(40)$, however the 
latter one is located in a less crowded spectral region.
From this reason McCartt \etal\ used the $P(40)$ line to produce calibration 
curve for concentration of radioactive carbon in the SCAR technique, where reference 
concentrations were determined by accelerator mass spectrometry. In their 
spectroscopic model, McCartt \etal\ use  a line intensity at 300~K taken from 
measurements by Galli \etal\ \cite{11GaBaBo.CO2} (9\% above UCL-IAO intensity) and 
line intensities of interfering isotopologues from the HITRAN2012 database. This 
leads to negative concentrations resulting from fitted calibration curve. One of 
the possible reason for that could be inaccurate line strength used in the 
retrieval model. Of equal importance are however: the accuracy of $^{14}$C  
abundance in samples and  the intensities  of the satellite lines of other carbon dioxide 
isotopologues. Here we provide line intensities which are internally consistent and proved to agree within experimental 
uncertainty to state-of-the-art measurements.
Table \ref{table:P20} also lists our prediction for the line 
intensity at $T = $ 170~K, a temperature which is commonly used for intensity 
measurements for the $P(20)$ line.

Vibrational assignments of the UCL line list for 646 were based on isotopic shifts of energy levels 
and respective assignments for the 626 and 636 isotopologues. After that, a comparison of the DVR3D line positions to experimental frequencies by Dobos \etal\  
\cite{89DoWiKl.CO2} was made. The tunable diode laser measurements supplied 
accuracy of 0.001 \cm\ or better in the 2229 -- 2259 \cm\ spectral range of the 
asymmetric stretching fundamental; this yields an RMSD $ = 0.004$ \cm. This result 
shows that calculated line positions deviate from experiment just above the 
stated experimental accuracy, hence may be considered as highly reliable for the 
00011 -- 00001 band. For this band  the average deviation from the EH calculations 
for 5 symmetric isotopologue levels is 0.018 \cm, which could probably be reduced by
treatment of mass-dependent non-Born-Oppenheimer effects.
A more recent study performed by Galli \etal\ \cite{11GaPaLo.CO2}, where high-resolution 
optical-frequency-comb-assisted cavity ring-down technique was used to measure 
ro-vibrational line positions in 2190 -- 2250 \cm\ region with accuracy of few MHz. 
Comparison with this study resulted in 0.005 \cm\ RMSD, thereby establishing the 
provisional uncertainty of the DVR3D line positions to 0.005 \cm\ for the 
asymmetric stretching fundamental. The study by Galli \etal\ awaits accurate intensity evaluation. This 
creates an opportunity for further utilization of our results and comparison 
with experiments, when done.

The 3$\nu_3$ family of bands discussed in ref. 
\cite{15ZaTePo.CO2} and in the preceding sections of this article, has been proven to be on 
average 12\% too weak in present calculations. For isotopologues other than the 
radioactive  $^{14}$C$^{16}$O$_2$ intensities for these bands were replaced with 
CDSD-296 entries. For $^{14}$C$^{16}$O$_2$ these intensities were scaled by 1.12 
to correct for the systematic differences known from other isotopologues.

\subsection{Recommended line lists}

For all six isotopologues two types of line lists were prepared. First, files named 
'UCL-296-isotopologue\_ name.dat' contain line positions calculated using Ames-1 PES with DVR3D 
program and line intensities using UCL DMS. Each line is supplemented with the
appropriate scatter factor, given in the last column (see supplementary 
materials). Vibrational assignments are taken from the newest version of the 
CDSD-296 database. The second set of line lists, called "recommended UCL-IAO 
line list" borrows line positions from the effective Hamiltonian calculations 
(see section \ref{line_positions}) whenever a match was possible between Ames-1 
PES based line positions from UCL and the effective Hamiltonian values. In rare 
cases when such match could not be found, line positions were transferred from 
the CDSD-296 database. The scheme used for the line intensities is similar the one
used for the main 
isotopologue. The $3 \nu_3$ bands and unstable lines were taken from the 
effective Hamiltonian calculations, and appropriate uncertainty indices were 
assigned. Intensities of stable lines belonging to bands stronger than 
$10^{-23}$ cm/molecule (for unit abundance) were taken from UCL DMS calculations 
and assigned HITRAN uncertainty code 8 (i.e. accuracy of 1\% or better). 
Stable lines belonging to parallel bands weaker than  $10^{-23}$ 
cm/molecule also come from UCL DMS computation and were given uncertainty code 7 (i.e. accuracy 1 -- 3 \%).
Intermediate lines and stable lines belonging to perpendicular bands weaker than $10^{-23}$ 
cm/molecule  feature HITRAN uncertainty code 6 (i.e. accuracy 3 -- 5\%). 
All line positions and line intensities for which 
scatter factor was not assigned were taken from CDSD-296. This was the case for 
only 3700 weak lines in total for all isotopologues. Both types of line lists are given in the 
supplementary materials with appropriate explanation in text files. Abundances 
were taken from the HITRAN2012 database with intensity cut-off $10^{-30}$ cm/molecule. For 
the radiocarbon isotopologue unit abundance was assumed and $10^{-27}$ 
cm/molecule intensity cut-off cm/molecule.

\section{Conclusion}

In the present study we compute new line lists for six symmetric isotopologues 
of carbon dioxide: $^{13}$C$^{16}$O$_2$, $^{14}$C$^{16}$O$_2$, 
$^{12}$C$^{17}$O$_2$, $^{12}$C$^{18}$O$_2$, $^{13}$C$^{17}$O$_2$ and 
$^{13}$C$^{18}$O$_2$. Detailed comparisons with both the theoretical and 
experimental works indicate the high accuracy of the line intensities resulting from our 
\abinitio\ DMS. Sensitivity analysis of the line intensities performed here 
confirmed the existence of several resonance interactions already reported in the 
literature, altogether with predictions for new pairs of energy levels involved 
in strong interactions. We hope that the methodology presented and evaluated here 
will be of use in theoretical approaches to other molecules. Accurate line 
positions generated using an effective Hamiltonian combined with highly accurate 
line intensities give comprehensive line lists, which we recommend for use in 
remote sensing studies and inclusions in databases.

This paper completes our analysis of the transition intensities of symmetric
isotopolgues of CO$_2$. We are currently analysing the transition intensities of the
asymmetric isotopologues. This work raises some theoretical issues as the loss
of symmetry has consequences for both the DVR3D nuclear motion calculations
and the CDSD effective Hamiltonian studies. Results will be reported in the near future
\cite{jtasym}.

\section*{Acknowledgments}
This work is supported by the UK Natural Environment Research Council (NERC) 
through grant NE/J010316,
the ERC under the Advanced Investigator Project
267219 and the Russian Fund for Fundamental Science.  The authors acknowledge 
the use of the UCL Legion High Performance Computing Facility (Legion@UCL), and 
associated support services, in the completion of this work.

\bibliographystyle{elsarticle-num}

\end{document}